\documentclass[journal]{IEEEtran}

\usepackage{amssymb,stmaryrd,amsmath,amsfonts,rotating}
\usepackage[noadjust]{cite}
\usepackage{color}
\usepackage[vflt]{floatflt}
\usepackage{epic}
\RequirePackage{bbm}
\allowdisplaybreaks
\newtheorem{theorem}{Theorem}
\newcommand{\btheo}{\begin{theorem}}
\newcommand{\etheo}{\end{theorem}}
\newcommand{\bproof}{\begin{proof}}
\newcommand{\eproof}{\end{proof}}
\newtheorem{definition}[theorem]{Definition}
\newcommand{\bdefi}{\begin{definition}}
\newcommand{\edefi}{\end{definition}}
\newtheorem{fact}[theorem]{Fact}
\newcommand{\bprop}{\begin{fact}}
\newcommand{\eprop}{\end{fact}}
\newtheorem{corollary}[theorem]{Corollary}
\newcommand{\bcor}{\begin{corollary}}
\newcommand{\ecor}{\end{corollary}}
\newtheorem{example}[theorem]{Example}
\newcommand{\bex}{\begin{example}}
\newcommand{\eex}{\end{example}}
\newtheorem{lemma}[theorem]{Lemma}
\newcommand{\blemma}{\begin{lemma}}
\newcommand{\elemma}{\end{lemma}}
\newtheorem{remark}[theorem]{Remark}
\newcommand{\bremark}{\begin{remark}}
\newcommand{\eremark}{\end{remark}}
\newtheorem{conj}[theorem]{Conjecture}
\newcommand{\bconj}{\begin{conj}}
\newcommand{\econj}{\end{conj}}

\newcommand{\reals}{\ensuremath{\mathbb{R}}}
\newcommand{\naturals}{\ensuremath{\mathbb{N}}}
\newcommand{\integers}{\ensuremath{\mathbb{Z}}}
\newcommand{\expectation}{\ensuremath{\mathbb{E}}}

\newcommand{\prob}{\ensuremath{\mathbb{P}}}

\def\0{{\tt 0}} 
\def\1{{\tt 1}} 
\def\?{{\tt *}} 
\newcommand{\graph}{{\ensuremath{\tt G}}}
\newcommand{\ldpc}{{{\ensuremath{\text{LDPC}}}}}
\newcommand{\ledge}{\ensuremath{\lambda}} 
\newcommand{\redge}{\ensuremath{\rho}} 
\newcommand{\Shasmall}{\ensuremath{\text{\tiny Sha}}} 
\newcommand{\GalBsmall}{\ensuremath{\text{\tiny GalB}}} 
\newcommand{\MPsmall}{\ensuremath{\text{\tiny MP}}} 
\newcommand{\iter}{\ell} %

\newcommand{\qed}{{\hfill \footnotesize $\blacksquare$}}
\renewcommand{\mid}{\,|\,}

\newcommand {\bE} {\mathbb{E}}
\newcommand {\gn} {\Omega}

\newcommand {\bad} {\mathcal{B}}

\newcommand {\nb} {\mathcal{N}}
\newcommand {\expn} {\mathcal{X}}
\newcommand{\errors}{{\ensuremath{\tt E}}}

\newcommand {\var} {\mathcal{V}}
\newcommand {\chset} {\mathcal{C}}

\newcommand {\msgspace} {\mathcal{M}}

\newcommand{\dr}{{\mathtt r}}
\newcommand{\dl}{{\mathtt l}}

\newcommand {\edge} {{\mathtt e}}
\newcommand {\vnode} {{\mathtt v}}
\newcommand {\cnode} {{\mathtt c}}

\newcommand {\M} {\mathcal{M}}

\newcommand{\E}{{\ensuremath{\tt E}}}

\newcommand{\LGalBsmall}{\ensuremath{\text{\tiny LGalB}}}

\newcommand{\LMSsmall}{\ensuremath{\text{\tiny LMS}}}
\newcommand{\MSsmall}{\ensuremath{\text{\tiny MS}}}

\newcommand{\EEx}{\hfill $\Diamond$}
\newcommand{\EDe}{\hfill $\Diamond$}

\newcommand {\ntouched} {{M}}
\newcommand {\witness} {\mathcal{W}}
\newcommand {\errorset} {\mathcal{E}}
\newcommand{\ind}{\mathbbm{1}}
\newcommand{\indicator}[1]{\ind_{\{ #1 \}}}

\newcommand{\expander}{\mathcal{X}}
\newcommand{\de}[1] {x_{#1}}

\begin{document}
\title{Exchange of Limits: Why Iterative Decoding Works}
\author{Satish Babu Korada and R{\"u}diger Urbanke\thanks{EPFL, School of Computer,
 \& Communication Sciences, Lausanne, CH-1015, Switzerland,
\{satish.korada, ruediger.urbanke\}@epfl.ch.}} %
\maketitle
\begin{abstract}
We consider communication over binary-input memoryless output-symmetric
channels using low-density parity-check codes and message-passing 
decoding.  The asymptotic (in the length) performance of such a
combination for a fixed number of iterations is given by density
evolution. Letting the number of iterations tend to infinity we get the density evolution
{\em threshold}, the largest channel parameter so that the bit error probability tends to 
zero as a function of the iterations.

In practice we often work with short codes and perform a large
number of iterations.  It is therefore interesting to consider what
happens if in the standard analysis we exchange the order in which
the blocklength and the number of iterations diverge to infinity.
In particular, we can ask whether both limits give the same threshold.

Although empirical observations strongly suggest that the exchange
of limits is valid for {\em all} channel parameters, we limit our discussion
to channel parameters below the density evolution threshold.
Specifically, we show that under some suitable technical conditions
the bit error probability vanishes below the density evolution
threshold regardless of how the limit is taken.
\end{abstract}


\section{Introduction}
\subsection{Motivation}\label{sec:motivation}
Consider transmission over a binary-input memoryless output-symmetric (BMS) channel
using a low-density parity-check (LDPC) code and decoding via a message-passing (MP) algorithm.
We refer the reader to \cite{RiU08} for an introduction to the standard notation and
an overview of the known results.
It is well known that, for good choices of the degree distribution
and the MP decoder, one can achieve rates close to the capacity of the channel
with low decoding complexity \cite{CFRU01}.

The standard analysis of iterative decoding systems assumes that the blocklength
is large (tending to infinity) and that a fixed number of iterations
is performed. As a consequence, when decoding a given bit, the output
of the decoder only depends on a fixed-sized local neighborhood of this bit 
and this local neighborhood is tree-like.
This local tree property implies that the messages arriving at nodes
are conditionally independent, significantly simplifying the analysis. To determine
the performance in this setting, we track the evolution of the message densities 
as a function of the iteration. This process is called {\em density evolution} (DE).
Denote the bit probability of error of a code $\graph$ 
after $\ell$ iterations by $P_b(\graph,\epsilon, \iter)$, where $\epsilon$ is the
channel parameter. Then DE computes 
\begin{equation}\label{lim1}
\lim_{n\to \infty}\bE[P_b(\graph,\epsilon, \iter)].
\end{equation}
If we now perform more and more iterations then we get a limiting performance 
corresponding to
\begin{equation}\label{lim11}
\lim_{\iter \to \infty}\lim_{n\to \infty}\bE[P_b(\graph,\epsilon,\iter)].
\end{equation}

In order for the computation graphs of depth $\ell$ to form a tree,
the number of iterations can not exceed $c \log(n)$, where $c$ is a constant that only depends on
the degree distribution. (For a $(\dl, \dr)$-regular degree distribution pair 
a valid choice of $c$ is $c(\dl, \dr)=\frac{2}{\log(\dl-1)(\dr-1)}$, \cite{Gal62}.)
In practice, this condition is rarely fulfilled: 
standard blocklengths measure only in the hundreds or thousands
but the number of iterations that have been observed to be useful in practice 
can easily exceed one hundred. 

Consider therefore the situation where we fix the blocklength but
let the number of iterations tend to infinity. This means, we consider
the limit
\begin{equation}\label{lim2}
\lim_{\iter\to \infty}\bE[P_b(\graph,\epsilon,\iter)].
\end{equation}
Now take the blocklength to infinity, i.e., consider
\begin{equation}\label{lim22}
\lim_{n\to\infty}\lim_{\iter\to \infty}\bE[P_b(\graph,\epsilon,\iter)].
\end{equation}
What can we say about (\ref{lim22}) and its relationship to (\ref{lim11})?  

Consider the belief propagation (BP) algorithm.
It was shown by McEliece, Rodemich, and Cheng \cite{MRC95} that one 
can construct specific graphs and noise realizations so that the messages on a specific
edge either show a chaotic behavior (as a function of iteration) or converge to limit cycles.
In particular, this means that the messages do not converge as a function of the iteration.
For a fixed length and a discrete channel, the number of graphs 
and noise realizations is finite. Therefore, if for single graph and noise realization the messages
do not converge as a function of $\ell$, then it is likely that
also $\lim_{\ell \rightarrow \infty} \bE[P_b(\graph,\epsilon,\iter)]$ does not
converge as a function of $n$ (unless by some miracle the various non-converging parts cancel).
Let us therefore consider 
$\limsup_{\ell \rightarrow \infty} \bE[P_b(\graph,\epsilon,\iter)]$ and
$\liminf_{\ell \rightarrow \infty} \bE[P_b(\graph,\epsilon,\iter)]$.
What happens if we increase the blocklength and consider
$\lim_{n \rightarrow \infty} \limsup_{\ell \rightarrow \infty} \bE[P_b(\graph,\epsilon,\iter)]$ and
$\lim_{n \rightarrow \infty} \liminf_{\ell \rightarrow \infty} \bE[P_b(\graph,\epsilon,\iter)]$?

We restrict our present study to the exchange of limits
{\em below the density threshold}. I.e., suppose that the given
combination (of the channel family and the MP decoder) has a threshold
in the following sense: for the given channel family characterized
by the real valued parameter $\epsilon$ there exists a threshold
$\epsilon^{\MPsmall}$ so that for all $0 \leq \epsilon <
\epsilon^{\MPsmall}$ the DE limit (\ref{lim11}) is $0$, whereas for
all $\epsilon > \epsilon^{\MPsmall}$ it is strictly positive.  
We will show that under suitable technical conditions the bit error
probability also tends to zero if we exchange the limits.  This
implies that the DE threshold is a meaningful and robust design parameter.

\subsection{Summary of Main Result}
Consider transmission over a BMS channel parametrized by $\epsilon$, using an $\ldpc{(n,\dl,\dr)}$
ensemble and decoding via an MP algorithm. Assume that the algorithm is symmetric
in the sense of \cite{RiU08}[Definition~4.81, p. 209]. 
Moreover, assume that this combination has a threshold and let $\epsilon^{\MPsmall}$ denote
this threshold.  If $\epsilon < \epsilon^{\MPsmall}$ then under the conditions stated in
Sections~\ref{sec:expansion} and \ref{sec:randomwalk},
\begin{align*}
\lim_{n\to\infty}\limsup_{\iter\to\infty}\bE[P_b^{\MPsmall}(\graph, \epsilon, \iter)]=0.
\end{align*}

Instead of considering just an exchange of limits one can consider
joint limits where the iteration is an arbitrary but increasing
function of the blocklength, i.e., one can consider $\lim_{n\to\infty}
\bE[P_b^{\MPsmall}(\graph, \epsilon, \iter(n))]$. Our
arguments extend to this case and one can show that 
\begin{align*}
\limsup_{n\to\infty} \bE[P_b^{\MPsmall}(\graph, \epsilon, \iter(n))]=0.
\end{align*}
But for the sake of simplicity we restrict
ourselves to the standard exchange of limits discussed above.
In the same spirit, although some of the techniques and statements we discuss extend
directly to the irregular case, in order to keep the exposition
simple we restrict our discussion to the standard regular ensemble
LDPC$(n, \dl, \dr)$.  

\subsection{Outline}
We introduce two techniques that are useful
in our context.  First, we consider expanders.  More precisely, in
Section~\ref{sec:expansion} we show that for codes with sufficient
expansion the exchange of limits is valid below the DE threshold.  The advantage
of using expansion is that the
argument applies to a wide variety of decoders. On the negative
side, the argument can only be applied to ensembles with large
variable-node degrees.

Why does expansion help in proving the desired result and why do
we need large variable-node degrees?  Assume that a sufficient number of
iterations has been performed so that the number of still erroneous
messages is relatively small.  Consider further iterations. There
are two reasons why a message emitted by a variable node can be
bad. This can be due to the received value, or it can be due to a
large number of bad incoming messages. If the degree of the variable
node is large then the received value becomes less and less important
(think of a node of degree $1000$ and a decoder with a finite number
of messages; in this case the received value has only a limited
influence on the outgoing message and this message is mostly
determined by the $999$ incoming messages).  If we ignore therefore
the received message then we see that expansion helps since it can
guarantee that only few nodes have many bad incoming messages;
otherwise the set of nodes that has bad outgoing messages
has too few neighbors in order for the graph to be an expander.

If the variable nodes have small degree, then the received values play
a significant role and can no longer be ignored.  Therefore, for small
degrees expansion arguments do not suffice by themselves.
In Section~\ref{sec:randomwalk} we concentrate on the case
$\dl=3$.  This is the smallest degree that is meaningful for all
the decoders that we consider and so one can think of it as the
most difficult general case.  Except for the
BEC, this case is not covered by a simple expansion argument and
the techniques are more involved.

\section{Sufficient Conditions Based on Expansion Arguments}
\label{sec:expansion}

Burshtein and Miller were the first to realize that expansion
arguments can be applied not only to the flipping algorithm but
also to show that certain MP algorithms have a fixed error correcting
radius \cite{BuM00}. Although their results can be
applied directly to our problem, we get stronger statements
by using the expansion in a slightly different manner.

\subsection{Definitions and Review}
\bdefi[Expansion] 
Let $\graph$ be an element from LDPC$(n, \dl,\dr)$.
~\\
1) Left Expander: The graph $\graph$ is an $(\dl,\dr,\alpha,\gamma)$ {\em left expander} 
if for every subset $\var$ of at most
$\alpha n$ variable nodes, the set of check nodes that are connected to $\var$
is at least $\gamma \vert\var\vert \dl $.
~\\
2) Right Expander: Let $m=n \frac{\dl}{\dr}$. The graph $\graph$ is an $(\dl,\dr,\alpha,\gamma)$ {\em right expander} if for every subset
$\chset$ of at most
$\alpha m$ check nodes, the set of variable nodes that are connected to
$\chset$
is at least $\gamma \vert\chset\vert \dr $.
\EDe
\edefi
Why are we using expansion arguments in the context of standard
LDPC ensembles? It is well known that such codes
are good expanders with high probability \cite{BuM00}. 
\begin{theorem}[Expansion of Random Graphs \cite{BuM00}]
\label{the:randomexpansion}
Let $\graph$ be chosen uniformly at random from LDPC$(n, \dl,\dr)$. 
Let $\alpha_{\max}$ be the positive solution of the equation 
\begin{align*}
\frac{\dl-1}{\dl}h_2(\alpha) - \frac{\dl}{\dr}h_2(\alpha \gamma \dr) -\alpha \gamma \dr h_2(1/\gamma \dr) = 0.
\end{align*}
Let $\expn(\dl,\dr,\alpha,\gamma)$ denote the set of graphs
\begin{align*}
\{\graph \in \ldpc(n,\dl,\dr): \text{ } \graph
\in (\dl,\dr,\alpha,\gamma) \text{ left expander}\}.
\end{align*}
If $\gamma < 1-\frac{1}{\dl}$ then $\alpha_{\max}$ is strictly positive and
for $\alpha < \alpha_{\max}$
\begin{align}\label{equ:expanderprob}
\prob\{\graph \in \expn(\dl,\dr,\alpha,\gamma) \} \geq 1-O(n^{-(\dl(1-\gamma) -1)}).
\end{align}
Let $m=n \frac{\dl}{\dr}$. We get the equivalent result for right expanders by exchanging the roles
of $\dl$ and $\dr$ as well as $n$ and $m$. 
\end{theorem}

As explained before, the idea is to show that the error probability goes to 
zero once the number of bad messages becomes smaller than a certain threshold. 
To make this more concrete we need a proper definition of ``good'' message subsets.
\bdefi[Good Message Subsets]\label{def:goodmsgset}
For a fixed $(\dl, \dr)$-regular ensemble and a fixed MP decoder with message
alphabet $\msgspace$,
let $\beta$,  $0 < \beta \leq 1$, be such that $\beta (\dl-1) \in
\naturals$.  A ``good'' pair of subsets of $\msgspace$ of ``strength''
$\beta$ is a pair of subsets $(G_\vnode,G_\cnode)$ so that
\begin{itemize}
\item
if at least $\beta(l-1)$ of the $(\dl-1)$ incoming messages at a variable node belong to
$G_\vnode$ then the outgoing message on the remaining {\em edge} is in $G_\cnode$
\item if all the $(\dr-1)$ incoming messages at a check node belong to
$G_\cnode$ then the outgoing message on the remaining {\em edge} is in $G_\vnode$
\item if at least $\beta(\dl-1) + 1$ of all $\dl$ incoming messages belong to $G_\vnode$, then the
{\em variable} is decoded correctly
\end{itemize}
We denote the probability of the bad message set $\msgspace\backslash G_\vnode$
after $\iter$ iterations of DE by $p_{\text{bad}}^{(\iter)}$.
\EDe
\edefi

As we will see shortly, for many MP decoders of interest the sets
$G_\vnode$ and $G_\cnode$ can be chosen to be equal. This is true for
all those MP decoders where the outgoing reliability at a check node is
equal to the least reliability of all the incoming messages (we call
them min-sum-type decoders).
Therefore, if all incoming messages are good (meaning they are
correct and have sufficiently large reliability) then the outgoing
message is correct and also has sufficiently large reliability.  
The BP decoder is an interesting case where $G_\vnode \neq G_\cnode$.  For this
decoder the reliability of the outgoing message at a check node is
{\em strictly} smaller than the smallest reliability of all incoming
messages.  Therefore, we need to
define the set $G_\cnode$ to consist of messages of strictly higher reliability
than the set of messages in $G_\vnode$.

\bdefi[Good Nodes]
We call a variable or check node ``good'' if 
all of its outgoing messages are good. All other nodes are called ``bad.''
\EDe
\edefi
%

\begin{example}[BEC and BP]\label{ex:BEC1}
If at least $1$ of the $(\dl-1)$ messages entering a variable node
is known then the outgoing message is known and if at least $1$ of
the $\dl$ messages entering a variable node is known then the
variable itself is known. Further, if all of the $(\dr-1)$ incoming
messages entering a check node are known then the outgoing message
is known. We conclude that {\em good} is equivalent to {\em known}
and that $\beta= \frac{1}{\dl-1}$.
\EEx
\end{example}

As a second standard example we consider transmission over the BSC$(\epsilon)$
and decoding via the so-called {\em Gallager Algorithm B} (GalB).
\bdefi[Gallager Algorithm B]
Messages are elements of $\{ \pm 1\}$.
The initial messages from the variable nodes to the check nodes are the
values received via the channel.
The decoding process
proceeds in iterations with the following processing rules:
\begin{itemize}
\item[] Check-Node Processing: At a check node the outgoing message along a
particular edge is
the product of the incoming messages along all the remaining edges.
\item[] Variable-Node Processing: At a variable node the outgoing message
along a particular edge is equal to the majority vote
on the set of other incoming messages and the received value. Ties are resolved
randomly.
\end{itemize}
\EDe
\edefi

\begin{example}[BSC and GalB]\label{ex:BSC1}
Assume that the received value (via the channel) is incorrect.
In this case
at least $\lceil(\dl-1)/2\rceil+1$ of the $(\dl-1)$ incoming messages should
be correct to ensure that the outgoing message is correct. If at least
$\lceil(\dl-1)/2\rceil+ 2$ of the $\dl$ incoming messages are correct
then the variable is decoded correctly. (In fact, it is sufficient to
have $\lfloor(\dl -1)/2 \rfloor + 2$ correct incoming messages to be able to
decode correctly.) 
Therefore, {\em good} is equivalent to {\em correct} and $\beta = \frac{\lceil (\dl-1)/2\rceil + 1}{\dl-1}$.
\EEx
\end{example}

\subsection{Expansion and Bit Error Probability}
\btheo[Expansion and Bit Error Probability]
\label{th:limexbit}
Consider an $\ldpc(n,\dl,\dr)$ ensemble, transmission over a BMS$(\epsilon)$ channel, 
and a symmetric MP decoder. 
Let $\beta$ be the strength of the good message subset. If $\beta <1 $ and
if for some $\epsilon$, $p_{\text{bad}}^{(\infty)} = 0$ then
\begin{align}
\lim_{n \to\infty}\limsup_{\iter\to
\infty}\bE_{\ldpc(n,\dl,\dr)}[P_b^{\MPsmall}(\graph,\epsilon,\iter)] = 0. \label{eq:limexbit}
\end{align} 
\etheo
\begin{proof}
Here is the idea of the proof: we first run the MP algorithm for a fixed number of iterations
such that the bit error probability is sufficiently small, say $p$. If the length $n$ is sufficiently
large then we can use DE to gage the number of required iterations.
Then, using the expansion properties of the graph, we show that the probability of error
stays close to $p$ for any number of further iterations. In particular, we show that
the error probability never exceeds $c p$, where $c$ is a constant, which only depends on
the degree distribution and $\beta$. Since $p$ can be chosen arbitrarily small, the claim follows.

Here is the fine print.  Define
\begin{align}\label{equ:defgamma}
\gamma = \Bigl(1-\frac{1}{\dl} \Bigr) \frac{1+\beta}{2} \stackrel{\beta < 1}{<}  \Bigl(1-\frac{1}{\dl} \Bigr).
\end{align}
Let $0 < \alpha < \alpha_{\max}(\gamma)$, where $\alpha_{\max}(\gamma)$ is the function defined in
Theorem~\ref{the:randomexpansion}.
Let $p = \frac{\alpha (1-\beta)(\dl-1)}{4}$ and let $\iter(p)$ be the number of 
iterations such that $p_{\text{bad}}^{(\ell)}\leq p$.
Since $p_{\text{bad}}^{(\infty)} = 0$ and $p > 0$ this is possible. 
Let $P_\edge(\graph,
\errors,\iter)$ denote the fraction of messages belonging to the bad set after
$\iter$ iterations. 
Let $\gn$ denote the space of code and noise realizations. Let $A \subseteq \gn$ denote the subset
\begin{align}
A = \{(\graph,\errors) \subseteq \gn \mid P_\edge(\graph, \errors,\iter(p)) \leq 2 p\} \label{def:setA}.
\end{align}
From (the Concentration) Theorem~\ref{the:concentration} we know that
\begin{align}
\label{equ:pofnotina}
\prob\{(\graph, \errors) \not \in A\} \leq 2 e^{-K n p^2}
\end{align}
for some strictly positive constant $K=K(\dl, \dr, p)$.  In words,
for most (sufficiently large) graphs and noise realizations the error probability after
a fixed number of iterations behaves close to the asymptotic ensemble.
We now show that once the
error probability is sufficiently small it never increases substantially
thereafter if the graph is an expander, regardless of how many
iterations we still perform.

Let $V_0 \subseteq [n]$ be the {\em initial} set of bad variable
nodes.  More precisely, $V_0$ is the set of all variable nodes that
are bad in the $\iter(p)$-th iteration.  We claim that $|V_0| \leq
\frac{2 p}{\dl-\beta(\dl-1)} n$. (This is because for a variable to send a bad
message it must have at least $\dl - \beta(\dl-1)$ incoming bad messages.) As we
just discussed, for most graphs and noise realizations
this is the case.  As a worst case we assume that all its outgoing
edges are bad. Let the set of check nodes connected to $V_0$ be
$C_0$. These are the only check nodes that potentially can send bad
messages in the next iteration. Therefore, we
call $C_0$ the initial set of {\em bad} check nodes.  Clearly,
\begin{align}\label{equ:trivialbound} |C_0| & \leq \dl|V_0|.
\end{align}

Consider a variable node and a fixed edge $\edge$ connected to it:
the outgoing message along $\edge$ is determined by the received
value as well as by the $(\dl-1)$ incoming messages along the other
$(\dl-1)$ edges. Recall that if $\beta(\dl-1)$ of those messages
are good then the outgoing message along edge $\edge$ is good.
Therefore, if a variable node has $\beta(\dl-1)+1$ good incoming
messages, then {\em all} outgoing messages are good. We conclude
that for a variable node to be bad at least $\dl-\beta(\dl-1)$
incoming messages must be bad.  Therefore, it should connect to at
least $\dl-\beta(\dl-1)$ bad check nodes. This leaves at most
$\beta(\dl-1)$ edges that are connected to {\em new} check nodes.

We want to count the number of bad variables that are created in
any of the future iterations. For convenience, once a variable
becomes bad we will consider it to be bad for all future iterations.
This implies that the set of bad variables is non-decreasing. 

Let us now bound the number of bad variable nodes by the following
process.  The process proceeds in discrete steps. At each step $t$,
consider the set of variables that are not contained in $V_t$ but
that are connected to at least $\dl-\beta(\dl-1)$ check nodes in
$C_t$ (the set of ``bad'' check nodes).  If at time $t$ no such
variable exists stop the process. Otherwise, choose one such variable
at random and add it to $V_t$. This gives us the set $V_{t+1}$.  We
also add all neighbors of this variable to $C_t$. This gives us the
set $C_{t+1}$.  By this we are adding the variable nodes that can
potentially become bad and the check nodes that can potentially
send bad messages to $V_t$ and $C_t$ respectively.  As discussed
above, for a good variable to become bad it must be connected to
at least $\dl - \beta(\dl-1)$ check nodes that are connected to
bad variable nodes.  Therefore, at most $ \beta(\dl-1)$ new
check nodes are added in each step. Hence, if the process continues
then 
\begin{align} |V_{t+1}|& =
|V_{t}| +1, \label{equ:variable} \\ |C_{t+1}| & \leq |C_t| +
\beta(\dl-1). \label{equ:check} \end{align} By assumption, the graph
is an element of $\expn(\dl,\dr,\alpha,\gamma)$.  Initially we have
$|V_0| \leq \frac{2 p}{\dl-\beta(\dl-1)} n = \frac{\alpha
(\dl-1)(1-\beta)}{2(\dl-\beta(\dl-1))} n\leq \alpha
n$.  Therefore, as long as $|V_t| \leq \alpha n$, \begin{align}
\label{equ:expansion} \gamma\dl |V_t| \leq |C_t|, \end{align} since
$C_t$ contains all neighbors of $V_t$.  Let $T$ denote the stopping
time of the process, i.e., the smallest time at which no new variable can
be added to $V_t$.  We will now show that the stopping time is
finite.  We have 
\begin{align*} \gamma\dl (|V_0|+t) &
\stackrel{\text{(\ref{equ:variable})}}{=} \gamma\dl |V_t|
\stackrel{\text{(\ref{equ:expansion})}}{\leq} |C_t|
\stackrel{\text{(\ref{equ:check})}}{\leq} |C_0| + t\beta(\dl-1) \\
& \stackrel{(\ref{equ:trivialbound})}{\leq}  \dl|V_0| + t\beta(\dl-1).
\end{align*} 
Solving for $t$ this gives us \begin{align*} T\leq
\frac{|V_0|\dl(1-\gamma)}{\gamma\dl -\beta(\dl-1)}.  \end{align*}
Therefore, \begin{align}\label{equ:cbound} |V_T| \leq
\frac{|V_0|\dl(1-\gamma)}{\gamma\dl -\beta(\dl-1)} +|V_0| \leq
\frac{2p}{\gamma \dl-\beta(\dl-1)}n = \alpha n, \end{align} where in the one before
last step we used the fact that $|V_0| \leq \frac{2p}{\dl - \beta(\dl-1)} n$.  The whole
derivation so far was based on the assumption that $|V_t| \leq
\alpha n$ for $0 \leq t \leq T$.  But as we can see from the above
equation, this condition is indeed verified ($|V_t|$ is non-decreasing
and $|V_T| \leq \alpha n$).

Putting all these things together, we get 
\begin{align}
\bE[P_b^{\MPsmall}(\graph, \epsilon, \iter)] 
= &  \bE[P_b^{\MPsmall}(\graph, \errors, \iter) (\indicator{(\graph,\errors) \in A}+
\indicator{(\graph,\errors) \not \in A})] \nonumber\\
\leq & \bE[P_b^{\MPsmall}(\graph, \errors, \iter) \indicator{(\graph,\errors) \in A}] + 
\prob\{(\graph, \errors) \not \in A\} \nonumber\\
\leq & \bE[P_b^{\MPsmall}(\graph, \errors, \iter) \indicator{(\graph,\errors) \in A}
\indicator{\graph \in \expn(\dl,\dr,\alpha,\gamma)}] + \nonumber \\
& \prob\{\graph \not \in \expn(\dl,\dr,\alpha,\gamma)\} + \prob\{(\graph, \errors) \not \in A\}. \nonumber
\end{align}
Apply $\limsup_{\iter \rightarrow \infty}$ on both sides of the inequality.
According to (\ref{equ:cbound}) the first term is bounded by $\alpha$. For the second
term, since $\gamma < 1-\frac{1}{\dl}$, we know from Theorem~\ref{the:randomexpansion}
 that it is upper bounded by $O(n^{-(\dl(1-\gamma) -1)})$. For the third term
we know from (\ref{equ:pofnotina})
that it is bounded by $2 e^{-K n p^2}$ for some strictly positive constant $K=K(\dl, \dr, p)$.
Therefore, if we subsequently apply the limit $\lim_{n \rightarrow \infty}$ then we get
\begin{align*}
\lim_{n\to\infty}\limsup_{\iter\to \infty}\bE[P_b^{\MPsmall}(\graph, \epsilon,
\iter)] & \leq \alpha.
\end{align*}
Since this conclusion is valid for any $0 < \alpha \leq \alpha_{\max}$ it follows that
\begin{align*}
\lim_{n\to\infty}\limsup_{\iter\to \infty}\bE[P_b^{\MPsmall}(\graph, \epsilon, \iter)] & = 0.
\end{align*}
\end{proof}

\begin{example}[BEC and BP]\label{ex:BEC2}
We know from Example~\ref{ex:BEC1} that $\beta(\dl-1) = 1$. If we apply
the conditions of Theorem~\ref{th:limexbit}, we see that we require $1/(\dl-1) < 1$. 
Hence, the exchange of the limits is valid for $\dl \geq 3$. Of course, for the
BEC the exchange of limits in this regime follows directly by the monotonicity of
the algorithm.
\EEx
\end{example}
\begin{example}[BSC and GalB]\label{ex:BSC2}
We know from Example~\ref{ex:BSC1} that $\beta(\dl-1) = \lceil (\dl-1)/2\rceil + 1$.
From Theorem \ref{th:limexbit} if $\epsilon < \epsilon^{\GalBsmall}$, the limits can be exchanged if  
$\dl-1 > 1 + {\lceil (\dl-1)/2\rceil}$, i.e., for $\dl \geq 5$. 
\EEx
\end{example}

The key to applying expansion arguments to decoders with a continuous
alphabet is to ensure that the received values are no longer dominant
once DE has reached small error probabilities. This can be achieved
by ensuring that the input alphabet is smaller than the message
alphabet.

\bdefi[Bounded MP Decoders]
Given a MP decoder whose message passing alphabet is unbounded, i.e., it
is equal to $\reals$, we associate to it a {\em bounded} version.
The {\em bounded} MP decoder with parameter $M \in \reals^+$, denote it by MP$(M)$,
is identical to the standard MP decoder
except that the reliability of the messages emitted by the check nodes is bounded to
$M$ before the messages are forwarded to the variable nodes.
\EDe \edefi
Note that the outgoing messages from the check nodes lie in $[-M,M]$
while the outgoing messages from the variable nodes can lie outside
this range.
\begin{example}[MS($M$), BP($M$) Decoders]
The MS$(M)$ decoder and the BP$(M)$ decoder are identical to the
standard min-sum (MS) and belief propagation (BP) decoder, except that the reliability of the
messages emitted by the check nodes is bounded to $M$ before the
messages are forwarded to the variable nodes.  \EEx \end{example}

\begin{example}[MS$(5)$ Decoder]
Consider an $(\dl \geq 5,\dr)$ ensemble and fix $M = 5$. Let the channel 
log-likelihoods belong to $[-1,1]$. It is easy to check that in this
case we can choose $G_\vnode = G_\cnode = [4,5]$ and that it has
strength $\beta \leq \frac34$. 
Therefore, if the probability of outgoing
messages from check nodes being in $[4,5]$ goes to $1$ under DE, then according
to Theorem~\ref{th:limexbit} the limits can be exchanged.

For example, consider BSC($\epsilon$) and $\ldpc(5,6)$ ensemble. It is known for
this channel and MS decoder the messages
are of the form $k \log\frac{1-\epsilon}{\epsilon}$, for $k\in \integers$.
Therefore we can restrict the message space to $\integers$ with the channel
values mapped to $\{\pm1\}$. 
Now, if we consider MS($5$) decoder, the messages belong to $\{-5,\dots,5\}$.
For this decoder, we can show that the limits can be exchanged till the DE threshold of $0.067$.
\EEx
\end{example}

\begin{example}[BP$(10)$ Decoder]
Let $\dl = 5$ and $\dr=6$ and fix $M=10$. Let the
channel log-likelihoods belong to $[-3,3]$. We claim that in  this case
the message subset pair $G_\vnode =[9,10], G_\cnode =[14,43]$ is good with
 strength $\beta = \frac34$.
This can be seen as follows: If all the incoming messages to a check node belong
to $G_\cnode$, then the outgoing message is at least $12.39$, which is mapped down to
$10$.
Suppose that at a variable node at least $3 (= \beta(\dl -1))$ out of the
$4$ incoming messages belong to $G_\vnode$. In this case the reliability of the
outgoing message is at least $14 = 3\times 9 - 10 -3$.
The maximum reliability is $43$. Moreover, if all the incoming messages
belong to $G_\vnode$ then the variable is decoded correctly. Therefore if the
probability of outgoing messages from check nodes being in $[9,10]$ goes to $1$
in the DE limit then from Theorem~\ref{th:limexbit}, the limits
can be exchanged. 

For example, consider BSC($\epsilon$) with channel log-likelihoods restricted 
between $[-3,3]$. For $\epsilon < \frac{1}{1+e^3}$, the log-likelihoods lie
outside $[-3,3]$ and hence they are mapped to $\{\pm 3\}$. In this case the
limits can be exchanged till the DE threshold of $0.136$. Note that this is what
is done practice, since one has to work with bounded likelihoods.
\EEx
\end{example}

\subsection{Expansion and Block Error Probability}
In the previous section we considered the bit error probability. We will now derive sufficient
conditions for the block error probability. Again we use expansion arguments but we proceed in
a slightly different way.
\btheo[Expansion and Block Error Probability]
\label{th:limexblock}
Consider an $\ldpc(n,\dl,\dr)$ ensemble, transmission over a BMS$(\epsilon)$ channel, 
and a symmetric MP decoder. 
Let $\beta$ be the strength of the good message subset. If $\beta <  \frac{\dl-2}{\dl-1}$ and
if for some $\epsilon$, $p_{\text{bad}}^{(\infty)} = 0$ then
\begin{align}
\lim_{n \to\infty}\limsup_{\iter\to
\infty}\bE_{\ldpc(n,\dl,\dr)}[P_B^{\MPsmall}(\graph,\epsilon,\iter)] = 0. \label{eq:limexblock}
\end{align} 
\etheo
\begin{proof}
As in Theorem \ref{th:limexbit} we first perform a fixed number of
iterations to bring down the bit error probability below a desired level.
We then use Theorem \ref{burshtein} to show
that for a graph with sufficient expansion the MP algorithm decodes the whole
block correctly once the bit error probability is sufficiently small.
This is very much in the spirit of Burshtein and Miller \cite{BuM00}.

Define 
\begin{align*}
\gamma =\left(1-\frac{1}{\dl}\right)\left(\frac{3+\beta}{4}\right).
\end{align*}
Let $0 < \alpha < \alpha_{\max}(\gamma)$, where $\alpha_{\max}(\gamma)$ is the function defined in
Theorem~\ref{the:randomexpansion}.
Let $p = \frac{\alpha(\dl-\beta(\dl-1))}{2 \dl\dr}$ and let $\iter(p)$ be the number of 
iterations such that $p_{\text{bad}}^{(\ell)} \leq p$.
Let $\gn$ denote the space of code and noise realizations. Let $P_\edge(\graph,
\errors,\iter)$ denote the fraction of messages belonging to the bad set after
$\iter$ iterations. Let $A \subseteq \gn$ denote the subset
\begin{align*}
A = \{(\graph,\errors) \subseteq \gn \mid P_\edge(\graph, \errors,\iter(p)) \leq 2 p\}.
\end{align*}
From (the Concentration) Theorem~\ref{the:concentration} we know that
\begin{align}
\label{equ:pofnotina2}
\prob\{(\graph, \errors) \not \in A\} \leq 2 e^{-K n p^2}
\end{align}
for some strictly positive constant $K=K(\dl, \dr, p)$. 

Since $\beta \frac{\dl-1}{\dl} \leq 2\gamma-1$ we can apply
Theorem~\ref{burshtein}: if $\graph \in \expn(\dl,\dr,\alpha,\gamma)$
and if the initial number of bad messages is less than $\frac{\alpha}{\dl\dr}$
then all the messages will become good
after a sufficient number of iterations. 

Putting all these things together, we get 
\begin{align}
\bE[P_B^{\MPsmall}(\graph, \epsilon, \iter)] 
= &  \bE[P_B^{\MPsmall}(\graph, \errors, \iter) (\indicator{(\graph,\errors) \in A}+
\indicator{(\graph,\errors) \not \in A})] \nonumber\\
\leq & \bE[P_B^{\MPsmall}(\graph, \errors, \iter) \indicator{(\graph,\errors) \in A}] + 
\prob\{(\graph, \errors) \not \in A\} \nonumber\\
\leq & \bE[P_B^{\MPsmall}(\graph, \errors, \iter) \indicator{(\graph,\errors) \in A}
\indicator{\graph \in \expn(\dl,\dr,\alpha,\gamma)}] + \nonumber \\
& \prob\{\graph \not \in \expn(\dl,\dr,\alpha,\gamma)\} + \prob\{(\graph, \errors) \not \in A\}. \nonumber
\end{align}
Apply $\limsup_{\iter \rightarrow \infty}$ on both sides of the inequality.
According to Theorem \ref{burshtein} the first term is $0$. For the second
term, since $\gamma < 1-\frac{1}{\dl}$, we know from Theorem~\ref{the:randomexpansion} that it is upper bounded by
$O(n^{-(\dl(1-\gamma) -1)})$. For the third term
we know from (\ref{equ:pofnotina2})
that it is bounded by $2 e^{-K n p^2}$ for some strictly positive constant
$K=K(\dl, \dr,p)$.
Therefore, if we subsequently apply the limit $\lim_{n \rightarrow \infty}$ then we get
\begin{align*}
\lim_{n\to\infty}\limsup_{\iter\to \infty}\bE[P_B^{\MPsmall}(\graph, \epsilon,
\iter)] & =0.
\end{align*}
\end{proof}

\begin{example}[BEC and BP]
According to Theorem~\ref{th:limexbit} we require $\dl \geq 4$. 
Hence, if $\dl \geq 4$ then the block error probability tends to zero
below the BP threshold.
\EEx
\end{example}

\begin{example}[BSC and GalB]
As explained in Example~\ref{ex:BSC1} for the Gallager B algorithm over BSC,
$\beta(\dl-1)=1+\lceil(\dl-1)/2\rceil$. The above condition implies if $\dl -2 >
1+\lceil(\dl-1)/2\rceil$, i.e., for $\dl\geq7$ the block error probability goes
to zero below $\epsilon^{\GalBsmall}$. 
\EEx
\end{example}
\begin{example}[MS$(5)$ Decoder]
Consider an $(\dl \geq 7,\dr)$ ensemble and fix $M = 5$. Let the channel 
log-likelihoods belong to $[-1,1]$. It is easy to check that in this
case we can choose $G_\vnode = G_\cnode = [4,5]$ and that it has
strength $\beta \leq \frac23$. Therefore, if the probability of outgoing
messages from check nodes being in $[4,5]$ goes to $1$ under DE
then according to Theorem~\ref{th:limexblock} the block error probability tends to $0$.
\EEx
\end{example}

\begin{example}[BP$(10)$ Decoder]
Let $\dl = 7$ and $\dr=8$ and fix $M=10$. Let the
channel log-likelihoods belong to $[-1,1]$. We claim that in  this case
the message subset pair $G_\vnode =[9,10], G_\cnode =[15,59]$ is good with
 strength $\beta = \frac23$. Therefore if the
probability of outgoing messages from check nodes being in $[9,10]$ goes to $1$
in the DE limit then from Theorem~\ref{th:limexblock}, the block error
probability goes to zero.
\EEx
\end{example}

Theorem~\ref{th:limexbit} has a stronger implication than
Theorem~\ref{th:limexblock} since it concerns the {\em block} error probability. 
Unfortunately, the required conditions are considerably more
restrictive. We conjecture that in fact the conditions of
Theorem~\ref{th:limexblock} can be weakened by considering several stages of
the algorithm jointly and that the required conditions are identical to
the ones in Theorem \ref{th:limexblock}. 
\bconj[Expansion and Block Error Probability]
\label{th:limexblock2}
Consider an $\ldpc(n,\dl,\dr)$ ensemble, transmission over a BMS$(\epsilon)$ channel, 
and a symmetric MP decoder. 
Let $\beta$ be the strength of the good message subset. If $\beta <  1$ and
if for some $\epsilon$, $p_{\text{bad}}^{(\infty)} = 0$ then
\begin{align}
\lim_{n \to\infty}\limsup_{\iter\to
\infty}\bE_{\ldpc(n,\dl,\dr)}[P_B^{\MPsmall}(\graph,\epsilon,\iter)] = 0.
\label{eq:limexblock2}
\end{align} 
\econj

\section{Sufficient Condition Based on Birth-Death Process}
\label{sec:randomwalk}
In the previous section we relied solely on the expansion of the
graph to prove the validity of the limit exchange. As can be seen
from the examples, for the decoders of interest the theorems are
only valid for higher degrees, lets say $\dl \geq 5$. 
Practical codes however typically have small degrees.  In these
cases expansion itself is not sufficient.  

In more detail, the proofs in the previous section have two phases.  In the first
phase we run the MP algorithm for some fixed number of iterations
to get the error probability down to a small constant.  In the
second phase we prove that the error probability stays close to $0$ regardless
of how many further iterations we perform and assuming pessimistically
that all variables nodes have bad received values.  This is too pessimistic 
an assumption for small degrees, where the received value plays an
important role.  In this section, we develop a method which takes
the actual channel realization into account.

Consider a MP decoder operating on a message alphabet $\msgspace\subseteq
\reals$.  Further, for $\mu \in \msgspace$, define $|\mu|$ to be
the {\em reliability} of the message. This means that we define the
reliability of $-\mu$ to be the same as the reliability of $\mu$.

Most of the MP algorithms used in practice like GalB, BP, and MS,
fall in the following category of {\em monotone} decoders.
\bdefi[Monotone MP Decoders] We say that a symmetric MP decoder is
monotone if the following conditions are fulfilled. At variable
nodes the processing rules are monotone with respect to the natural
order on $\msgspace$; for a fixed received value, the outgoing
message is a non-decreasing function of the incoming messages.

At check nodes the processing rules are monotone with respect to
the natural order on the reliabilities;  the reliability of the
outgoing message is a non-decreasing function of the reliabilities
of the incoming messages.  \EDe \edefi
Monotonicity is a useful property and it is also quite natural.
A remaining difficulty in analyzing these decoders is that at check
nodes the monotonicity is with respect to the reliability and not
the message itself. We will see shortly how to get around this problem.

In what follows we mainly discuss the case of the GalB algorithm
and $\dl=3$.  The generalization to degree $\dl\geq4$ is straightforward
and it is discussed in Section~\ref{sec:extensions}. In this section we further
give some examples of other monotone decoders to which the method
can be extended.

\subsection{Main Result and Outline}
\blemma[Exchange of Limits]\label{lem:limex3}
Consider transmission over the BSC($\epsilon$) using random elements from the 
$(\dl, \dr)$-regular ensemble and decoding by the GalB algorithm. If $\epsilon < \epsilon^{\LGalBsmall}$ then
\begin{align*}
\lim_{n\to\infty}\limsup_{\iter\to\infty}\bE[P_b^{\GalBsmall}(\graph, \epsilon,\iter)] = 0,
\end{align*}
where  $\epsilon^{\LGalBsmall}$ is the smallest parameter $\epsilon$ 
for which a solution to the following fixed point equation 
exists in $(0,\epsilon]$.
\begin{align}\label{eqn:LGalBFP}
x & = \epsilon\sum_{k=0} ^{\lfloor {\frac{\dl-1}{2}}\rfloor} {\dl-1\choose
k}y^{k} (1-y)^{\dl-1-k }\nonumber\\
& + \bar{\epsilon} \sum_{k=\lfloor{{\frac{\dl}{2}}}\rfloor +1 } ^{\dl-1} {\dl-1 \choose
k}(1-y)^k y^{\dl-1-k }\nonumber \\
& + \frac{\indicator{\frac\dl2 \in \naturals}}{2}  {\dl-1\choose
\frac\dl2}\Big( \epsilon y^{\frac\dl2} (1-y)^{\frac\dl2-1 } +
\bar{\epsilon} (1-y)^{\frac\dl2} (y)^{\frac\dl2-1}\Big),
\end{align}
where $y = (1-x)^{\dr-1}$.
For the case of ($\dl=3,\dr$)-regular ensemble this equation simplifies to
\begin{align*}
x & = \bar{\epsilon}(1-(1-x)^{\dr-1})^{2}+\epsilon (1-(1-x)^{2 (\dr-1)}).
\end{align*}
\elemma
Discussion: Note that the threshold $\epsilon^{\LGalBsmall}$
introduced in the preceding lemma is in general slightly smaller than
the DE threshold $\epsilon^{\GalBsmall}$.
We pose the extension of the result to channel
values up to the DE threshold as an interesting open problem.
It is likely to be difficult.

\begin{table}[!h]
\begin{center}
\begin{tabular}{r|l|l|l|l}
$\dr$ & $\text{rate}$ & \phantom{xx}$\epsilon^{\Shasmall}$ &  \phantom{xx}$\epsilon^{\GalBsmall}$ & \phantom{xx}$\epsilon^{\LGalBsmall}$  \\ \hline
$3$ & $0.0$ & $\approx 0.5$ &  $\approx 0.222$ & $\approx 0.1705$  \\
$4$ & $0.25$ & $\approx 0.2145$ &  $\approx 0.1068$ & $\approx 0.0847$  \\
$5$ & $0.4$ & $\approx 0.1461$ &  $\approx 0.06119$ & $\approx 0.0506$  \\
$6$ & $0.5$ & $\approx 0.11002$ &  $\approx 0.0394$ & $\approx 0.0336$  \\
$7$ & $0.5714$ & $\approx 0.08766$ &  $\approx 0.02751$ & $\approx 0.02398$  \\
$8$ & $0.625$ & $\approx 0.07245$ &  $\approx 0.02027$ & $\approx 0.01795$  \\
$9$ & $0.667$ & $\approx 0.06141$ &  $\approx 0.01554$ & $\approx 0.01395$  \\
$10$ & $0.7$ & $\approx 0.05324$ &  $\approx 0.01229$ & $\approx 0.01115$  \\
\end{tabular}
\caption{\label{tab:thresholds3} Threshold values for some degree distributions
with $\dl= 3$.}
\end{center}
\end{table}

\begin{table}[!h]
\begin{center}
\begin{tabular}{r|l|l|l|l}
$\dr$ & $\text{rate}$ & \phantom{xx}$\epsilon^{\Shasmall}$ &  \phantom{xx}$\epsilon^{\GalBsmall}$ & \phantom{xx}$\epsilon^{\LGalBsmall}$  \\ \hline
$4$ & $0.0$ & $\approx 0.5$ &  $\approx 0.0840$ & $\approx 0.0697$  \\
$5$ & $0.2$ & $\approx 0.1461$ &  $\approx 0.0464$ & $\approx 0.0399$  \\
$6$ & $0.333$ & $\approx 0.11002$ &  $\approx 0.0292$ & $\approx 0.0258$  \\
$7$ & $0.4286$ & $\approx 0.08766$ &  $\approx 0.0200$ & $\approx 0.018$  \\
$8$ & $0.5$ & $\approx 0.07245$ &  $\approx 0.0146$ & $\approx 0.0133$  \\
$9$ & $0.556$ & $\approx 0.06141$ &  $\approx 0.0111$ & $\approx 0.0102$  \\
$10$ & $0.6$ & $\approx 0.05324$ &  $\approx 0.0087$ & $\approx 0.0081$  \\
\end{tabular}
\caption{\label{tab:thresholds4} Threshold values for some degree distributions with $\dl = 4$.}
\end{center}
\end{table}

\begin{example}
Table~\ref{tab:thresholds3} shows thresholds for $\dl=3$, $\dr=3, \cdots, 10$.
For the $(\dl=3, \dr=6)$ degree distribution we have  $\epsilon^{\LGalBsmall} \approx 0.0336$.
This is slightly smaller than, but comparable to, $\epsilon^{\GalBsmall} \approx 0.0394$.
\EEx
\end{example}

We proceed by a sequence of simplifications, ensuring
in each step that the modified algorithm is an upper bound on the original process.
In Section~\ref{sec:lgb} we simplify the decoder by ``linearizing'' the processing rules
at the check nodes.
In Section~\ref{sec:marking} we further upper bound the process by considering the marking process
associated with the decoding algorithm.
In Section~\ref{sec:witness} we construct a witness for the marking process
and derive bounds on the size of such a witness.
In Section~\ref{sec:randomization} we then show that, conditioned on the witness, we 
can consider the channel realizations outside the witness to be random and independent of the witness.
In Section~\ref{sec:backtoexpansion} we use an expansion argument to bound the stopping time
of the birth and death process associated with the marking process. 
Finally, in Section~\ref{sec:puttingitalltogether} we combine all previous statements
to derive at our conclusion.

\subsection{Linearized Gallager Algorithm B}\label{sec:lgb}
We proceed as in Section~\ref{sec:expansion}: Fix $0 \leq \epsilon < \epsilon^{\LGalBsmall}$. We 
prove that for every $\alpha>0$ there exists an $n(\alpha, \epsilon)$ so that
$\limsup_{\iter\to\infty} \bE[P_b^{\GalBsmall}(\graph, \epsilon, \iter)]<\alpha$ for 
$n \geq n(\alpha, \epsilon)$.

Without loss of generality we can assume that the all-one codeword was sent.
We will make this assumption throughout the remainder of this section.
Therefore, the message $1$ signifies in the sequel a {\em correct} message,
whereas $-1$ implies that the message is {\em incorrect}.

For this setting, we define the following {\em linearized} version of the decoder.  
\bdefi[Linearized GalB]\label{def:LGalB} The {\em linearized} GalB
decoder, denoted by LGalB, is defined as follows: at the variable
node the computation rule is same as that of the GalB decoder. At
the check node the outgoing message is the minimum of the incoming
messages.  \edefi 
Discussion: The LGalB is not a practical decoding algorithm but
rather a convenient device for analysis;  
it is
understood that we assume that the all-one codeword was transmitted
and that quantities like the error probability refer to the 
variables decoded as $-1$.
By some abuse
of notation, we nevertheless refer to it as a decoder. 

The LGalB decoder is monotone also with respect to the incoming messages at
check nodes. Moreover, it satisfies the following property.
\begin{lemma}[LGalB is Upper Bound on GalB] 	 
\label{lem:LGalBUB} 	 
For any graph $\graph$, any noise realization $\errors$, 	 
any starting set of ``bad'' edges, and any $\iter$, 	 
we have $P^{\GalBsmall}_\edge(\graph, \errors, \iter) \leq
P^{\LGalBsmall}_\edge(\graph, \errors, \iter)$, 	 
where $P_\edge(\graph, \errors, \iter)$ denotes the fraction of
erroneous messages after 	 
$\iter$ iterations of decoding. 	 
\end{lemma} 	 
\begin{proof} 	 
Consider one iteration, i.e., a check-node step followed by a
variable-node step. Let $\bad_{\ell}^{\GalBsmall/\LGalBsmall}$ 
denote the set of bad edges (edges with message $-1$) after the
$\ell$-th iteration of GalB and LGalB, respectively. Let
$\psi^{\GalBsmall/\LGalBsmall}_{\errors}(\bad)$ denote the set of
bad edges after one iteration assuming that the initial 
such set is $\bad$.

We use the following two facts: (i) The outgoing messages for the
LGalB decoder at variable/check nodes are monotone; if we decrease
(with respect to the natural order on $\msgspace$) the input at a
variable/check node then the output is either decreased or stays the same. I.e.,
if $\bad \subseteq \bad'$, meaning that the messages
in $\bad'$ can be obtained by decreasing some of the $+1$ messages in $\bad$
to $-1$, then $\psi_{\errors}^{\LGalBsmall}(\bad)
\subseteq \psi_{\errors}^{\LGalBsmall}(\bad')$.  (ii) For any set
of input messages, the outgoing message of LGalB is less than or
equal to the message of the GalB decoder, i.e.,
$\psi_{\errors}^{\GalBsmall}(\bad)
\subseteq\psi_{\errors}^{\LGalBsmall}(\bad) $.

For the proof, we proceed by induction. Let $\bad_0$ be the initial
set of bad edges. After the first iteration, from (ii) we get
$\bad_1^{\GalBsmall}
=\psi_{\errors}^{\GalBsmall}(\bad_0)\subseteq\psi_{\errors}^{\LGalBsmall}(\bad_0)=\bad_1^{\LGalBsmall}$.
 To complete the proof it is sufficient to show
that $\bad_\ell^\GalBsmall \subseteq \bad_\ell^{\LGalBsmall}$ implies
$\bad_{\ell+1}^\GalBsmall \subseteq \bad_{\ell+1}^{\LGalBsmall}$.
Using (i) and (ii) we have $\bad_{\ell+1}^{\LGalBsmall} =
\psi_{\errors}^{\LGalBsmall}(\bad_\ell^{\LGalBsmall}) \supseteq
\psi_{\errors}^{\LGalBsmall}(\bad_\ell^{\GalBsmall}) \supseteq
\psi_{\errors}^{\GalBsmall}(\bad_\ell^\GalBsmall) = \bad_{\ell+1}^{\GalBsmall}$
and hence the lemma.  \end{proof}

From the above lemma it suffices to prove the exchange of limits
for the linearized algorithm. Note that $\epsilon^{\LGalBsmall}$
as defined in Lemma~\ref{lem:limex3} is the threshold of the LGalB
algorithm. We will prove that for every $0 \leq \epsilon <
\epsilon^{\LGalBsmall}$ and every $\alpha>0$ there exists an
$n(\alpha, \epsilon)$ so that $\limsup_{\iter\to\infty}
\bE[P_b^{\LGalBsmall}(\graph, \epsilon, \iter)]<\alpha$ for $n \geq
n(\alpha, \epsilon)$.  As we will see later, the monotonicity
property of LGalB considerably simplifies the analysis. But the
price paid for the simplification is that the technique works only
for $\epsilon <\epsilon^{\LGalBsmall}$, which is slightly smaller
than the DE threshold.

\subsection{Marking Process}\label{sec:marking}
Rather than analyzing the LGalB algorithm directly, we analyze the
associated {\em marking process}.  This process is monotone as
a function of the iterations.

More precisely, we split the process into two phases:
we start with LGalB for $\iter(p)$ iterations to 
get the error probability below $p$;
we then continue the marking process associated with an infinite number
of further iterations of LGalB. This means that we mark any variable that is
bad in at least one iteration $\ell \geq \ell(p)$. 
Clearly, the union of all variables
that are bad at at least one point in time $\ell \geq \ell(p)$ is an upper bound
on the maximum number of variables that are bad at any specific 
instance in time.

The standard {\em schedule} of the LGalB is parallel, i.e., all
incoming messages (at either variable or check nodes) are processed
at the same time.  This is the natural schedule for an actual
implementation. For the purpose of analysis it is convenient to
consider an {\em asynchronous} schedule.

Here is how the general asynchronous marking process proceeds.  We
are given a graph $\graph$ and a noise realization $\errors$.  We
are also given a set of {\em marked} edges. These  marked edges are
directed, from variable node to check node. At the start of the process
mark the variable nodes that are connected to the marked edges.  Declare
all other variables and edges as {\em unmarked}. Unmarked edges do
not have a direction. The process proceeds in discrete steps. At
each step we pick a marked edge and we perform the processing described
below. We continue until no more marked edges are left. Here are the processing rules:

\noindent If the marked edge $\edge$ goes from variable to check: 
\begin{itemize}
\item Let $\cnode$ be the check node connected to $\edge$. 
Declare $\edge$ to be {\em unmarked} but {\em mark} all other edges
connected to $\cnode$; orient these marked edges from check to variable;
\end{itemize}
If the marked edge $\edge$ goes from check to variable:
\begin{itemize}
\item Let $\vnode$ be the connected variable node. 
If $\vnode$ has a {\em good} associated channel realization and $\vnode$ is unmarked
then mark $\vnode$
and declare $\edge$ to be unmarked.
\item  Let $\vnode$ be the connected variable node. If $\vnode$ 
has an associated {\em bad} channel realization or if $\vnode$ 
has an associated {\em good} channel realization but
is {\em marked}:  (i) mark $\vnode$ and all its outgoing edges; (ii)
orient the edges from variable to check; (iii) unmark $\edge$.
\end{itemize}

Let $\M(\graph,\errors, {\mathcal S})$ denote the set of 
marked variables assuming that we start with the set of marked edges ${\mathcal S}$
and that we run the asynchronous marking process.
Let $M(\graph,\errors, {\mathcal S})  = |\M(\graph,\errors,{\mathcal S})|$.
As a special case, let $\M(\graph,\errors, \iter)$ denote the set of marked
variables at the end of the process assuming that the initial set of marked
edges is the set of bad edges after $\ell$ rounds of LGalB. As before,
$M(\graph,\errors, \iter)  = |\M(\graph,\errors,\iter)|$. 

It is not hard to see that for any $\ell' \geq \ell$,
$P_b^{\LGalBsmall}(\graph, \epsilon, \ell') \leq M(\graph,\errors, \iter)/n$:
for $\ell'=\ell$ both processes start with the same set of bad edges and both
are operating on the same graph and noise realization.
At the check-node side the processing rules are identical. At the variable-node side 
both processes also behave
in the same way if they encounter a variable node with a bad channel realization.
The difference lies in the behavior when they encounter a variable node
with a good channel realization. In such a case the outgoing message for
the LGalB is bad only if there are two bad messages entering {\em at the same
time instance}. The asynchronous marking process algorithm declares the outgoing message to 
be bad if there are two incoming bad messages, even if the two messages might
correspond to different time instances as measured by the parallel schedule.
We conclude that for $\ell' \in \naturals$
\begin{align}\label{equ:LGalBversusM}
\limsup_{\iter\to\infty}\bE[P_b^{\LGalBsmall}(\graph, \epsilon, \ell)] 
\leq \frac{1}{n} \bE[M(\graph,\errors, \iter') ].
\end{align}

\subsection{Witness}\label{sec:witness}
It remains to bound $\bE[M(\graph,\errors,\iter)]$. 
Assume at first that we take a random graph $\graph$ and a random noise realization $\errors$
and that we start the marking process with a sufficiently small {\em random} set of marked edges 
(and not the set of bad edges after $\iter$ iterations of LGalB).
In this case one can show that the number of marked nodes at the end of the process
is with high probability not more than a constant multiple of the size of 
the starting set. To prove this statement, we use the fact that
the graph, the noise, and the starting set of edges are all independent.
Therefore, the marking process behaves essentially like a birth and death process: we pick an edge and
we explore its neighborhood; with a certain probability the edge dies (if it enters
a variable node with a correctly received value) and with a certain
probability the edge spawns some children. As long as the expected number
of new children is less than $1$ the process eventually dies with probability $1$.

Unfortunately our situation is more involved. After $\iter$ iterations
the starting set of marked edges is correlated, both with the graph
as well as with the noise realization. Our aim therefore is to
reduce this correlated case to the uncorrelated case by a sequence
of transformations.  As a first step we show how to get rid of the
correlation with respect to the noise realization.

Consider a fixed graph $\graph$. Assume that we have performed
$\iter$ iterations of LGalB.  For each edge $\edge$ that is bad in the
$\ell$-th iteration we construct a ``witness."  A witness for 
$\edge$ is a subset of the computation tree of height $\iter$ (where height is
counted as the number of variable node levels)
for $\edge$ consisting of paths that carried bad messages in the past iterations. We construct
the witness recursively starting with $\edge$. Orient $\edge$ from check node
to variable node. At any point in time while constructing the witness
associated with $\edge$ we have a partial witness that is a tree with oriented edges.
The initial such partial witness is $\edge$. One step in the construction
consists of taking a leaf edge of the partial witness and to  
``grow it out'' according to the following rules.  

\begin{figure}[htp]
\setlength{\unitlength}{0.65bp}%
\begin{picture}(500,180)(45,40)
\put(0,0){\includegraphics[scale=0.65]{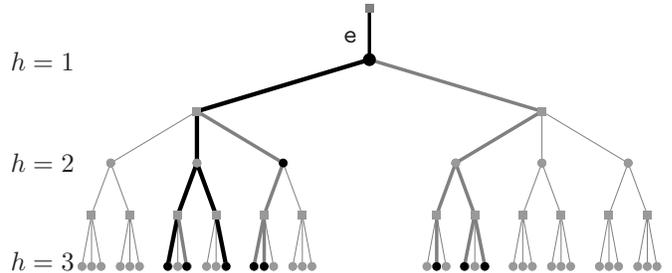}}
\put(235,160){$\edge$}
\put(42,144){$h =1$}
\put(42,85){$h =2$}
\put(42,28){$h =3$}
\end{picture}
\caption{\label{fig:witness} Construction of the witness for a bad edge $\edge$. The
{\em dark} variables represent channel errors. The part of the tree with {\em
dark} edges represent the witness, the {\em thick} edges, including both dark and
gray, represent the bad messages in the past iterations. The number $h$ in the
left indicates the height of the tree.}
\end{figure}

If an edge enters a
variable node that has an incorrect received value then add the
{\em smallest} (according to some fixed but arbitrary order on the
set of edges) edge that carries an incorrect incoming message to
the witness and continue the process along this edge. The added
edge is directed from variable node to check node.  If an edge
enters a variable node that has a correct received value then add
both incoming edges to the witness and follow the process along
both edges. (Note that in this case both of these edges must have
carried bad messages.) Again, both of these edges are directed from variable
to check node.  If an edge enters a check node then choose the
smallest incoming edge that carries an incorrect message and add
it to the witness.  Continue the process along this edge.  The added
edge is directed from check to variable node.  Continue the process
until depth $\ell$.  Fig.~\ref{fig:witness}
shows an example for $\dl=3$, $\dr=4$,  and $\ell=3$.

Denote the union of all witnesses for all edges that are bad in
the $\ell$-th iteration by  $\witness(\graph,\errors, \ell)$.  We
simply call it {\em the witness}. The witness is a part of the
graph that on its own explains why the set of bad edges after
$\iter$ iterations is bad.

How large is $\witness$?  The larger $\ell$, the fewer bad edges we
expect to see in iteration $\ell$.  On the other hand, the size of
the witness for each bad edge grows as a function of $\ell$.  The
next lemma, whose proof can be found in Appendix~\ref{appen:sizeofwitness},
asserts that the first effect dominates and that the expected size of
$\witness$ converges to zero as the number of iterations increases.
\blemma[Size of Witness]\label{lem:sizeofwitness} Consider the
$(3,\dr)$-regular ensemble.  For $0 \leq \epsilon<\epsilon^{\LGalBsmall}$,
\begin{align*} \lim_{n \rightarrow \infty} \frac{1}{n}\bE[|\witness(\graph,
\errors, \iter)|] = o_\iter(1).  \end{align*} \elemma

Why do we construct a witness?  It is
intuitive that if we keep the witness fixed but randomize the
structure as well as the received values on the remainder of the
graph then the situation should only get worse: already the witness
itself explains all the bad messages and hence any further bad channel values can
only create more bad messages. In the next two sections we
show that under some suitable technical conditions this intuition is indeed
correct.

\subsection{Randomization}\label{sec:randomization}
A witness $\witness$ consists of two parts, (i) the graph structure of
$\witness$ and (ii) the channel realizations of the variables in $\witness$. 
We will often need to refer to either of these parts on their own.
By some abuse of notation we write $\witness$ also if we refer only to the graph
structure or only to the channel realizations.
The usage should be clear from the context.
As an example, we write
$\witness \subseteq \graph$ to indicate that $\graph$ contains $\witness$ as a subgraph
and we write $\witness \subseteq \errors$ to indicate that the received values of all variables
in $\witness$ agree with the values that these variables take on in $\errors$. 

Fix a graph $\graph$ and a witness $\witness$, $\witness \subseteq \graph$. 
Let $\errorset_{\graph,\witness}$ denote the set of all error realizations
$\errors$ that give rise to $\witness$, i.e.,  $\witness(\graph,
\errors,\ell)=\witness$. Clearly, for all $\errors\in\errorset_{\graph,\witness}$ we must have
$\witness \subseteq \errors$. In words, on the set of variables fixed by the witness
the errors are fixed by the witness itself. 
Therefore, the various $\errors$ that create
this witness differ only on $\graph\backslash\witness$. 
As a convention, we define  
$\errorset_{\graph,\witness} =\emptyset$ if $\witness \not \subseteq \graph$.

Let  $\errorset'_{\graph, \witness}$ denote the set of projections of
$\errorset_{\graph,\witness}$ onto the variables in $\graph\backslash \witness$.
Let $\errors' \in \errorset'_{\graph,\witness}$. Think of
$\errors'$ as an element of $\{0, 1\}^{|\graph\backslash\witness|}$,
where $0$ denotes a correct received value and $1$
denotes an incorrect received value. In this way, $\errorset'_{\graph, \witness}$ is
a subset of $\{0,1\}^{|\graph\backslash\witness|}$.

This is important: $\errorset'_{\graph, \witness}$ has structure.
We claim that,
if $\errors' \in \errorset'_{\graph, \witness}$ then $\errorset'_{\graph, \witness}$ also contains
$\errors'_{\leq}$ (as defined in Appendix~\ref{appen:FKG}). More precisely, if the noise realization
$\errors' \in \errorset'_{\graph, \witness}$ gives rise to the witness $\witness$ then
converting any incorrect received value in $\errors'$ to a correct one will
also give rise to $\witness$. This is true since the LGalB algorithm is monotone, so that
taking away some incorrectly received values can not increase the size of bad edges
observed in the $\ell$-th iteration. But on the other hand, $\witness$ itself ensures that the
set of bad edges after $\ell$ iterations includes all the bad edges we saw originally.
The proof of the following lemma relies heavily on this property.

\begin{lemma}[Channel Randomization]\label{lem:randomization1}
Fix $\graph$ and let $\witness \subseteq \graph$.
Let $\bE_{\errors'}[\cdot]$ denote the expectation with respect to the channel realizations
$\errors'$ in $\graph\backslash\witness$.
Then 
\begin{align}\label{eq:randomization}
& 
\bE_{\errors'}
[
\ntouched(\graph, (\witness,\errors'), \witness)
\indicator{\errors'\in \errorset'_{\graph, \witness}}] \nonumber\\
& \phantom{xxxxxx}\leq
\bE_{\errors'}
[\ntouched(\graph, (\witness, \errors'), \witness) ]
\bE_{\errors'}[\indicator{\errors' \in\errorset'_{\graph, \witness}}].
\end{align}
\end{lemma}
Discussion: Lemma~\ref{lem:randomization1} has the following important operational significance.
If we divide both sides by $\bE_{\errors'}[\indicator{\errors' \in\errorset'_{\graph, \witness}}]$,
the left-hand side is the expectation of marked variables, where the expectation
is computed over all those channel realizations that give rise to the given witness $\witness$,
whereas the right-hand side gives the expectation over all channel
realizations (outside the witness) regardless whether they give rise to $\witness$ or not.
Clearly, the right-hand side is much easier to compute, since the channel is now
independent of $\witness$. The lemma states that, if we assume that the channel
outside $\witness$ is independently chosen then we get an upper bound on the size of the marked
variables.

\begin{proof}
Let $n'=|\graph\backslash\witness|$. Let $P\{ \cdot \}$ be the probability measure associated with
$\bE_{\errors'}[ \cdot ]$, i.e., $P\{\E'\} =
\epsilon^{n_1}\bar{\epsilon}^{n'-n_1}$, where $n_1$ denotes the number of ones in $\E'$. Let $f(\E')$
denote the function $\indicator{\E'\in \errorset'_{\graph,\witness}}$,
and let $g(\E')$ denote the function $M(\graph,(\witness, \errors'),
\witness)$. Note that $f$ is a decreasing function on
$\{0,1\}^{n'}$ because if $f(\E') = 1$ then
for all $\E''\leq \E'$, $f(\E'') = 1$. Further, $g$ is an increasing
in $\{0,1\}^{n'}$ since LGalB is monotone in the
number of channel errors. Since $g(\E')\leq n$,
$n-g$ is non-negative and it is a decreasing function.  For $s,t
\in \{0,1\}^{n'}$, let $|s|$ denote the
number of $1$s in $s$ and $s\vee t$ and $s\wedge t$ be as defined in
Appendix \ref{appen:FKG}. Then, 
\begin{align*} 
P\{s\}P\{t\}& = \epsilon^{|s|+|t|}(\bar{\epsilon})^{n'-|s|-|t|}, \\ 
P\{s\vee t\}& = \epsilon^{|s|+|t|-|s \wedge t|}(\bar{\epsilon})^{n'-(|s|+|t|-|s \wedge t|)},
\\ P\{s\wedge t\}& = \epsilon^{|s\wedge t|}(\bar{\epsilon})^{n'-|s\wedge t|}. 
\end{align*} 
Therefore, $P\{s\} P\{t\} = P\{s\vee t\}P\{s\wedge t\}$. 
Applying the FKG inequality in the form of Lemma \ref{lem:fkg} to $f$ and $n-g$, we get 
\begin{align*} 
\bE[f(n-g)] & \geq \bE[f]\bE[n-g].
\end{align*} 
This implies $\bE[fg] \leq \bE[f]\bE[g]$. 
\end{proof}

We can now upper bound the right-hand side of (\ref{equ:LGalBversusM}).
The proof of the next lemma can be found in Appendix~\ref{apen:randomization}.
\begin{lemma}[Markov Inequality]\label{lem:randomization2}
Consider the $(\dl=3, \dr)$-regular ensemble and transmission over the BSC$(\epsilon)$.
Let $(\graph, \errors)$ be chosen uniformly at random. 
Let $\iter \in \naturals$ and $\theta>0$ so that 
$\bE[|\witness(\graph,\errors,\iter)|] \leq \theta^2 n$.
Then
\begin{align*} 
& \bE[\ntouched(\graph,\errors,\iter)] \\
& \leq \sum_{\witness : |\witness| \leq {\theta}n}
\sum_{\graph}
\prob\{\graph\}
 \prob\{\errorset_{\graph,\witness}\} 
\bE_{\errors'}
[\ntouched(\graph, (\witness, \errors'), \witness) ]  +{\theta}n .
\end{align*}
\end{lemma}
\subsection{Back to Expansion}\label{sec:backtoexpansion}
In the previous section we have shown that for a fixed graph $\graph$, and a
given witness $\witness$, we can ignore the correlations between the witness
and the {\em channel values} in  $\graph\backslash\witness$ and consider those channel values to be chosen
independently. But the {\em graph structure} of $\graph\backslash\witness$ is still correlated
with $\witness$. Let us now deal with this correlation and get a bound
on the marking process for those $\graph$ that have an expansion close to the typical one of the ensemble.

Consider the following random process, which we call the R-process. The process
proceeds in discrete steps
and has {\em state} $(C_t, S_t, B_t, I_t)$  at {\em time} $t$, 
where each component is an integer.
We initialize the process with $(C_0, S_0, B_0, I_0)=(0,S_0, 0, 0)$,
where $S_0 \in \naturals$.

At each step we have two choices.
We can either perform a {\em regular} step or a {\em boundary} step.
The effect of each step type on the state $(C_t, S_t, B_t, I_t)$ is shown
in Table~\ref{tab:transitions}. If we choose a regular step then,
with probability $\epsilon$, an {\em extension} step is executed and, with
probability $\bar{\epsilon}$, a {\em pruning} step is performed. 
The choices of extension step versus pruning step are iid.

\begin{table}[!h]
\begin{center}
\begin{tabular}{l|r|r|r|r} 
& $C_t$ & $S_t$ & $B_t$ & $I_t$ \\ \hline
\text{regular extend} & $\mathbf{2}$ & $\mathbf{2\dr - 3}$ & $\mathbf{0}$ & $\mathbf{1}$ \\
& $1$ & $\dr-3$ & $0$ & $1$ \\
& $0$ & $-3$ & $0$ & $1$ \\
\text{regular prune} & $\mathbf{0}$ & $\mathbf{-1}$ & $\mathbf{1}$ & $\mathbf{0}$ \\
\text{boundary} & $\mathbf{1}$ & $\mathbf{\dr-2}$ & $\mathbf{-1}$ & $\mathbf{1}$ \\
& $0$ & $-2$ & $-1$ & $1$  
\end{tabular}
\caption{\label{tab:transitions} Possible state transitions.
Note that there are several possible transitions corresponding to a ``regular extend'' step
as well as a ``boundary'' step.
As explained below, the transitions indicated in bold letters dominate the other transitions
in the sense of Definition~\ref{def:ordering}.}
\end{center}
\end{table}

In our choice of step type we are restricted by the following:
at any time during the process the state has to satisfy 
\begin{align}\label{eq:expansion1}
\gamma \dr C_t  \leq S_t + B_t + I_t,
\end{align} 
where $\gamma = 1-\frac{1+\delta}{\dr}$ for some strictly positive number
$\delta$.
Let $T$ be the smallest time $t$ so that $S_t=0$.
It is convenient to formally define the process for all $t$ by
setting $U_t=U_T$ for $t \geq T$.

Discussion: Here is the interpretation of the above process.  We
are given a fixed graph $\graph$ and a witness $\witness$. The 
channel realizations in $\graph\backslash\witness$ are generated independently with
probability of error $\epsilon$. We are interested in computing the expected
number of marked variables
$\bE_{\errors'}[\ntouched(\graph,(\witness,\errors'),\witness)]$.

The components of the
state vector have the following interpretation.
By some further abuse of notation, 
let $\witness$ refer now also to the variables contained in $\witness$.
Let $\nb(\witness)$ denote all the check nodes that neighbor
$\witness$. 
We start our process with those edges connected to $\nb(\witness)$
that do not connect to $\witness$. The cardinality of this set
is denoted by $S_0$ (where the ``s" stands for {\em surviving}).  
In each step
we take a single edge from this set of
surviving edges and ``grow it out.'' 

Let us discuss this process in more detail. When we ``grow out'' an edge
we first visit the connected variable node. 
Suppose that this is the first time that
the process visits this variable node. We call this a {\em regular}
step.

If the received value of this variable node is good then we stop the process along this edge.
We add the variable to the {\em boundary} set to make a mental note that we have
seen this node exactly once.
The  boundary set has cardinality $B_{t}$. 
We further subtract $1$ from $S_t$ to take into account that we finished processing
one of the ``surviving'' edges.

If the received value is bad then we add this variable node to the {\em internal}
variable nodes. The cardinality of this set is $I_{t}$. 
This means that in this step we increase $I_t$ by $1$.
Further,
we expand the graph along the two outgoing edges, add the (at most) two
connected check nodes to the set of internal check nodes (whose cardinality
is denoted by $C_t$) and add all the remaining edges that emanate from 
these check nodes to the set of surviving edges. This adds (at most) $2(\dr-1)$ new
survivors, but we have to subtract the edge we started from. Therefore, $S_t$ is
increased by at most $2\dr-3$.

So far we have assumed that we have not seen the variable node
(that is connected to the edge which we grow out) before.
Suppose now that, to the contrary, the variable is an element of the boundary.
We know that in this case the received value is good, but we also know that the
variable received another bad incoming message. Therefore, the variable will
send a bad outgoing message along its remaining edge. Hence,
we move this variable node from the boundary to the internal set (this decreases $B_t$ by
$1$ and increases $I_t$ by $1$).  Further,
we grow out the graph along the only remaining outgoing edge. This 
adds at most one new check node and at most $\dr-1$ outgoing edges to 
the set of surviving edges. Discounting again the edge we started with, 
we add in total at most  $\dr-2$ to $S_t$. 

Suppose that the graph $\graph$ is a right expander; 
i.e., $\graph \in \expn(\dl,\dr,\alpha,\gamma)$,
where $\gamma \geq 1-\frac{1+\delta}{\dr}$ for some strictly positive $\delta$.  This means that every
collection ${\cal C}$ of check nodes of size at most $\alpha m$
has at least $\gamma |{\cal C}| \dr$ connected variable nodes.  Consider
the state of the system at some time $t$. At this point in time we
have $C_t$ check nodes. All these check nodes are ``internal,'' i.e., all their
neighboring variable nodes are either counted in $V_t$ or $I_t$, or they are yet
to be encountered by the process which cannot be more than the survivors set $S_t$. We know that $\graph$ is an expander
and suppose for now that $C_t \leq \alpha m$. Then we know that the
number of connected variable neighbors must be at least $\gamma \dr
C_t$, i.e., at any time during the process the state should satisfy
\begin{align}
\gamma \dr C_t & \leq S_t +B_t + I_t. \label{eqn:expansion}
\end{align}
We claim that
\begin{align}
\gamma \dr C_{t} & \leq S_{t}+B_{t} + I_{t} - (1-\delta)\label{eqn:necessary}
\end{align}
is a necessary condition to be able to perform a boundary step at time $t$.
To see this, suppose we take a boundary step. If you look at Table~\ref{tab:transitions}
you will see that there are two possible transitions. One can check that the
transition stated in bold letters gives the less restrictive condition. 
Let us therefore only focus on this case.
The state after applying the boundary state must still fulfill (\ref{eqn:expansion}).
This means that we must have 
\begin{align*}
\gamma \dr (C_{t}+1) & \leq (S_{t}+\dr-2)+(B_{t}-1) + (I_{t}+1).
\end{align*} 
The claim is proved by rewriting this inequality.

From the above discussion we claim that for a given $\witness$ and $\graph$, where
$\graph \in \expn(\dl,\dr,\alpha,\gamma)$, as long as $C_t\leq \alpha m$ then the
marking process can be modeled as the R-process. 
The random variable $I_\infty$ is equal to the random variable
$\ntouched(\graph,(\witness,\errors'),\witness) - |\witness|$ of the marking process (we subtract the
size of witness because we do not include it in the internal variables).
For the actual marking process the decision of whether a regular step or a boundary step is
taken is forced by the structure of the graph and our choice of which edge to grow out.
For the R-process the role of graph is taken by a {\em strategy}. A strategy is
any (randomized) decision function $F$ that, based on the initial state and past decisions and outcomes, 
decides whether a regular step or a boundary step is taken at any point in time.

Here is the connection between the actual physical process and the R-process in
more detail. Assume we are given a graph $\graph$ and a witness $\witness$. 
We know the graph and therefore we also know which edges of the graph are elements
of the surviving set. Therefore, when we pick a survivor, we know in advance
whether the step is a regular step or a boundary step. The noise realization, which 
is not known to us a priori, determines whether a regular step is a regular extend or prune step.
We see that each graph gives rise to a strategy. As long as
the size of all revealed nodes
is sufficiently small this strategy will be admissible since the expansion will be valid
up to this point.

Since we are only interested in an upper bound
on the number of marked variables, we allow the R-process to use an arbitrary {\em strategy}, only
limited by the condition (\ref{eqn:expansion}). 
We call a strategy which obeys (\ref{eqn:expansion}) an {\em admissible} strategy. 
Since the actual physical process
is also limited by (\ref{eqn:expansion}) (under the condition that the graph is
an expander and the process has not grown beyond the size where
the expansion is valid), 
it suffices to derive upper bounds on $\expectation[I_{\infty}]$ that 
is valid for all choices of the strategy.

We relax one further restriction imposed by the actual physical process in 
order to simplify our task. Again, this only increases $\expectation[I_{\infty}]$. 
In the marking process, we can only perform a
boundary step if the boundary set is strictly positive. In other words, we require $B_t > 0$ for a boundary
step to be performed. We lift this restriction for the R-process.

\bdefi[Ordering of States] \label{def:ordering}
The state $U\equiv (C,S,B,I)$ dominates the state $U'\equiv(C',S',B',I')$, denoted by
$U\geq U'$ if 
\begin{itemize}
\item[(i)]$S\geq S'$, 
\item[(ii)] $I\geq I'$,
\item[(iii)] $S+B+I- \gamma \dr C \geq  S'+B'+I'-\gamma\dr C'$.
\end{itemize}
\EDe
\edefi
\blemma[Monotonicity of $I_\infty$ with State]\label{lem:domination}
Consider the R-process with admissible strategy $F'$ and initial state
$U'\equiv(C',S',B',I')$. Let $U\equiv(C,S,B,I)$ be an initial state which dominates $U'$, i.e.,
$U\geq U'$. Then there exists an admissible strategy $F$ so that 
$\bE[I_\infty(U, F)] \geq \bE[I_\infty(U', F')]$, 
where $I_\infty(U, F)$ denotes $I_\infty$ assuming that the R-process is
initialized with $U$ and that the process uses the strategy $F$.
\elemma
\begin{proof}
Given $U'$ and the admissible strategy $F'$ we construct the
admissible strategy $F$ in the following way.
The process with initial state $U$ uses strategy $F'$ but
applies it to the {\em pseudo} state $U'$. 
Further, it updates its pseudo state
according to the realization of the process and bases its future decisions on strategy $F'$ applied
to this evolving pseudo state. Call the phase of the process until the pseudo state
has reached $S'=0$ the ``initial'' phase of the process.
At that point the $(U, F)$ process switches to any admissible strategy based on its real state. To be concrete,
assume that it uses a {\em greedy strategy} at this point. This means that the process 
performs a boundary step any time it is admissible.

In order to show the desired inequality on the expected values we 
couple the processes $(U', F')$ and $(U, F)$. We imagine that we run both
processes in parallel and that they
experience exactly the same randomness
(this refers to the randomness contained in the choice of the 
transitions as well as any randomness which might be used by the strategy).
Assume for the moment that strategy $F$ is admissible. 

In the initial phase of the algorithm (until the $(U', F')$ process stops because $S_t'=0$)
the $(U, F)$ process proceeds in lock-step with the $(U', S')$ process. 
Since $S_0 \geq S_0'$ and since $S_t-S_0=S_t'-S_0'$ it
follows that $S_t \geq S_t'$ in this initial phase. This means that the process
$(U, F)$ never stops before the process $(U', F')$. Further, $I_0 \geq I_0'$, 
$I_t-I_0=I_t'-I_0'$, and $I_t$ is a non-decreasing function.
It follows that for every realization $I_{\infty}(U, F) \geq I_{\infty}(U', F')$.
This implies, {\em a fortiori}, the claimed inequality on the expected values.

Let us now show that the protocol $F$ is admissible. We claim that for all $t \in \naturals$ 
\begin{align} \label{equ:admissible} 
S_t+B_t+I_t- \gamma \dr C_t \geq  S_t'+B_t'+I_t'-\gamma\dr C_t'.
\end{align}
By definition this is true for $t=0$. But by construction of the coupling,
$S_t-S_0=S_t'-S_0'$,  $I_t-I_0=I_t'-I_0'$,  $B_t-B_0=B_t'-B_0'$,  and
$C_t-C_0=C_t'-C_0'$. It follows that the left-hand side in (\ref{equ:admissible})
is always at least as large as the right-hand side. Therefore, if $F'$ is admissible then
so is $F$.
\end{proof}

From Table~\ref{tab:transitions} we see that for regular extend and boundary
steps there are several possible outcomes. For each of these two steps,
there is a single outcome (highlighted in the table) whose resulting state
dominates those of the other outcomes. Since we are interested in an upper
bound on $I_\infty$, thanks to the above lemma, we can restrict our attention to these dominating steps.  

Consider the {\em greedy} strategy, call if $F^g$. For this greedy strategy,
whenever (\ref{eqn:necessary}) is true we perform a
boundary step. 
\begin{lemma}[Domination of the Greedy Process] For a given initial state
$U=(C_0,S_0,B_0,I_0)$ and any admissible strategy $F$, we have 
\begin{align*}
\bE[I_\infty(U, F^g)]\geq \bE[I_\infty(U, F)].
\end{align*}
\end{lemma}
\begin{proof}
Again we construct a coupling between the processes $(U, F)$ and $(U, F^g)$. 
As remarked above, for both processes we can assume that the state transitions
are the ones indicated in bold in Table~\ref{tab:transitions}. The only randomness
therefore resides in whether for a regular step the process {\em extends} or {\em prunes}
and, possibly, in the randomness used for the strategy $F$. There is no randomness
involved in any boundary steps. The coupling consists in coupling for each regular step $i$, $i \in \naturals$,
the outcomes of these regular steps. 
In more detail, if for the process $(U, F)$ the $i$-th regular step
results in a pruning then the same occurs for the $i$-th regular step for
the process $(U, F^g)$. By construction, for all regular steps
the change of $S$, $I$, $B$, and $C$ is the same for both processes.
Assume we measure ``time'' not in the absolute number of steps taken
but by the number of regular steps taken. Consider a process $(U, F)$ and assume
that this process is still ``alive'' at `time $t$.
Then its state $U_t$ only depends on the realization
of the random variables during the regular steps and on the total number of boundary
steps taken, but it does not depend on the order of the steps taken. 

Since the process $(U, F^g)$ has by definition done at least as many boundary
steps as the process $(U, F)$ it further follows that if we compare the two processes
at ``time'' $i$ corresponding to $i$ regular steps then the number of survivors (and also
the number of internal nodes)
for $(U, F^g)$  is at least as large as the number of survivors for $(U, F)$.
Therefore, if at this time the process $(U, F)$ is still alive then so is the
process $(U, F^g)$ and the latter has at least as many accumulated internal variable nodes
as the former. This proves our claim.
\end{proof}

Since we are interested in {\em upper} bounding $\bE[I_\infty]$, it is sufficient
to bound $\bE[I_\infty(U, F^g)]$, which is done in the next lemma. We use large
deviation properties of the sub-critical Galton-Watson process. For the convenience
of the reader we provide this estimate in Appendix \ref{appen:birthdeath}.

\blemma[Birth Death Process]\label{lem:birthdeath}
Let the initial state be $U=(0,S_0,0,0)$. Fix a strictly positive $\delta$, $0 < \delta < \frac{1}{2(\dr-1)}$,
so that $\frac{1-\delta}{2 \delta} \in \naturals$ and let $\gamma =
1-\frac{1+\delta}{\dr}$. 
For all $\epsilon < \frac{1}{2(\dr-1)}$ there
exist constants $c = c(\dl,\dr,\epsilon,\delta)$, $c > 1$, and $c'=c'(\dl,\dr,\epsilon,\delta) >0$
so that
\begin{align*}
\prob\{ I_\infty(U, F^g) \geq c S_0 \} \leq e^{-c'S_0}.
\end{align*}
\elemma
\begin{proof}
Since condition (\ref{eqn:necessary}) is satisfied in the beginning, the greedy
R-process starts with some boundary steps. We claim that after exactly  
$\lfloor \frac{S_0}{1-\delta} \rfloor$ such boundary
steps the condition (\ref{eqn:necessary}) is for the first time no longer fulfilled.
To see this, ignore the integer constraint for a moment.
At the beginning of the process the condition
(\ref{eqn:necessary}) reads $0 \leq S_0-(1-\delta)$. After
$\frac{S_0}{1-\delta}$ boundary steps
this condition is transformed to
\begin{align*}
\gamma \dr \frac{S_0}{1-\delta} \leq S_0 + \frac{S_0}{1-\delta} (\dr-2) -(1-\delta),
\end{align*}
which is equivalent to $0 \leq -(1-\delta)$. We see that the inequality is no longer fulfilled
and it is easy to check that this is the first time that it is no longer fulfilled.

After the initial boundary steps, the greedy strategy performs regular steps
until exactly $\frac{1-\delta}{2\delta}$ regular extend steps are performed and
then follows it by exactly one boundary step. This sequence is then repeated.
(Note that by our assumption $\frac{1-\delta}{2\delta} \in \naturals$.)

To see this, note that each regular extend step increases the right-hand side of 
(\ref{eqn:expansion}) by $2(\dr-1)$ and the left-hand side by
$2(\dr-1-\delta)$. Further, each boundary step increases the left-hand side by
$\dr-1 -\delta$ and the right-hand side by $\dr-2$. Since
$\frac{1-\delta}{2\delta} 2(\dr-1-\delta) + (\dr-1-\delta) = 
\frac{1-\delta}{2\delta} 2(\dr-1) + (\dr-2)$, we see that after one such sequence of first 
$\frac{1-\delta}{2\delta}$
regular extends steps
followed by a boundary step the inequality is unchanged (up to an added constant).
(A regular prune step does not change the condition (\ref{eqn:expansion}).)

Since the randomness is contained only in the regular steps,
we can model the process as consisting of only regular steps. To include the
effect of boundary steps, we alter the outcome of the regular extend step as
follows. From Table~\ref{tab:transitions} note that for each regular
extend step we increase $S$ by $2\dr-3$ and $I$ by $1$. We include the effect of
boundary step by changing this to an increment of
$2\dr-3+(\dr-2)\frac{2\delta}{1-\delta}$  for $S$ and
$1+\frac{2\delta}{1-\delta}$ for $I$, respectively. 

Now this process is a standard birth and death process. Recall that we have
$\epsilon<\frac{1}{2(\dr-1)}$ and $\delta < \frac{1}{2(\dr-1)}$. 
Hence, the expected increase in $S$ at each step is 
$\epsilon (2\dr-3+(\dr-2)\frac{2\delta}{1-\delta})$. This is strictly less than $1$.
As discussed in more detail in Appendix~\ref{appen:birthdeath}, this shows that,
except for an exponentially small probability, this process stops for
$t\leq cS_0$ for some appropriate constant $c > 1$. This proves our lemma since in each step we
create at most $1+\frac{2\delta}{1-\delta}$ internal variables.
\end{proof}

Using Lemma~\ref{lem:birthdeath} we bound the
number of variables marked by the marking process as follows.

\blemma[Upper Bound]\label{lem:upperbound}
Let $\gamma =1-\frac{1+\delta}{\dr}$ for some $0 < \delta < \frac{1}{2(\dr-1)}$.
Fix $\graph$ and $\witness$ such that $\witness\subseteq\graph$ and $\graph \in
\expn(\dl,\dr,\alpha,\gamma)$. Let $c = c(\dl,\dr,\epsilon,\delta)$ be the constant
appearing in Lemma~\ref{lem:birthdeath}.
If $|\witness| \leq \frac\dl{6c(\dr-1)\dr} \alpha n$ then
\begin{align*}
\lim_{n\to\infty}\frac{1}{n}\bE_{\errors'}[\ntouched(\graph,(\witness,\errors'),\witness)] & \leq
\alpha \frac{\dl}{\dr}.  
\end{align*}
\elemma 
\begin{proof} 
Let $m=\frac{\dl}{\dr} n$.  
The maximum number of surviving edges coming out of the witness $\witness$ is
$3(\dr-1)|\witness|$. Let this be $S_0$. 
Consider the R-process with initial state $U=(0,S_0,0,0)$
and the greedy strategy $F^g$.
From Lemma~\ref{lem:birthdeath} there exists a strictly positive constant $c'$ such that 
\begin{align*}
\prob\{I_{\infty}(U, F^g) \geq c S_0 \} \leq e^{-c'S_0}.
\end{align*} 
The bound on $|\witness|$ in the hypothesis implies that $c S_0 = c 3(\dr-1)|\witness| \leq
\frac{\alpha}{2} m$. From Table~\ref{tab:transitions} we
see that any time the number of internal variable nodes is increased by $1$
the number of check nodes increases by at most $2$.
Therefore, $I_{\infty}(U, F^g) \leq c S_0$ implies that $C_\infty \leq 2cS_0\leq \alpha m$.
This shows that the expansion property is satisfied for the whole duration of the process. 
Hence, $I_\infty(U, F^g)$ is a valid upper bound for $M(\graph,(\witness,\errors'),\witness)$.

Let $M(\errors')$ denote $M(\graph,(\witness,\errors'),\witness)$. Since 
$M(\errors')$ counts the initial $|\witness| < \frac{\alpha}{2}m$ variables present in 
$\witness$ along with the internal
variables created,
\begin{align*}
\prob\{M(\errors')\geq \alpha m\} \leq \prob\{I_\infty(U, F^g) \geq \frac{\alpha}{2} m\} \leq e^{-c'S_0}.
\end{align*}
Therefore, 
\begin{align*}
\bE_{\errors'}[&\ntouched(\graph,(\witness,\errors'),\witness)] \\
&\leq \prob\{M(\errors') \leq \alpha m \} \alpha m + \prob\{M(\errors') \geq
\alpha m \} n \\
& \leq \alpha \frac{\dl}{\dr}n + (1-e^{-\frac{c'\alpha \dl n}{2\dr}})n.
\end{align*}
The lemma is proved by taking the limit $n\to\infty$.
\end{proof}

\subsection{Putting It All Together}\label{sec:puttingitalltogether}
In this section we prove Lemma~\ref{lem:limex3} using the results developed in
the previous sections.

\noindent{\em Proof of Lemma~\ref{lem:limex3}.}
Recall that we consider an $(\dl=3, \dr)$-regular ensemble and
that $0 \leq \epsilon < \epsilon^{\LGalBsmall}$.

Fix $0< \delta < \frac{1}{2(\dr-1)}$ and define $\gamma = 1-\frac{1+\delta}{\dr}$. 
Let $\alpha_\text{max}(\gamma)$ be the constant defined in Theorem~\ref{the:randomexpansion}.
Note that  $\alpha_\text{max}(\gamma)$ is strictly positive since $\delta$ is strictly positive.

Choose $0 < \alpha < \alpha_\text{max}(\gamma)$.
Let $\expn(\dl,\dr,\alpha,\gamma)$ denote the set of graphs
$\{\graph \in \ldpc(n,\dl,\dr): \text{ } \graph
\in (\dl,\dr,\alpha,\gamma) \text{ right expander}\}$.
From Theorem~\ref{the:randomexpansion} we know that
\begin{align}\label{eqn:probofexpansion}
\prob\{\graph\not\in\expn\} = o_n(1).
\end{align}

Let $c=c(\dl,\dr,\epsilon,\delta)$ be the coefficient appearing in 
Lemma ~\ref{lem:birthdeath} and
define $\theta = \frac{\dl}{6c(\dr-1)\dr} \alpha$. 
From Lemma~\ref{lem:sizeofwitness} we know that there exists an iteration 
$\ell$ such that
\begin{align}
\lim_{n\to\infty}\frac{1}{n}\bE[|\witness(\graph,\errors,\ell)|] \leq \frac12 \theta^2.
\end{align}
Let $n(\theta)$ be such that for $n\geq n(\theta)$,
$\bE[|\witness(\graph,\errors,\ell)|] \leq \theta^2 n$.

Using Lemma \ref{lem:randomization2}, and splitting the expectation over $\expn$
and its complement, we get

\begin{align*}
& 
\bE[
\ntouched(\graph, \errors, \iter)
] \\
&\leq \sum_{\witness:|\witness| \leq {\theta}n}
\sum_{\graph:\graph\in \expn}
\prob\{\graph\} \prob\{\errorset_{\graph, \witness}\}
\bE_{\errors'}
[\ntouched(\graph, (\witness, \errors'), \witness)] +\\
&\;\;\;\;\sum_{\witness:|\witness| \leq {\theta}n}
\sum_{\graph:\graph\not\in \expn}
\prob\{\graph\} \prob\{\errorset_{\graph, \witness}\}
\bE_{\errors'}
[\ntouched(\graph, (\witness, \errors'), \witness)]+\\ 
&\;\;\;\;\;{\theta}n.
\end{align*}
Consider the first term. From Lemma~\ref{lem:upperbound} we know that
\begin{align}\label{eqn:firstterm}
\bE_{\errors'}[M(\graph,(\witness,\errors'),\witness)] 
\leq \alpha\frac{\dl}{\dr} n + o(n). 
\end{align}
Consider the second term. Bound the expectation by $n$ and remove the restriction
on the size of the witness. This gives the bound 
\begin{align*}
&
 \sum_{\witness} \sum_{\graph:\graph\not \in \expn}
\prob\{\graph\}\prob\{\errorset_{\graph, \witness}\} n.
\end{align*}
Switch the two summations and use the fact that, for a given $\graph$,
each $\errors$ realization maps to only one $\witness$. We get 
\begin{align}
\sum_{\graph:\graph\not\in\expander}
\prob\{\graph\}\sum_{\witness:\witness\subseteq\graph}
\prob\{\errorset_{\graph, \witness}\}  & = \sum_{\graph:\graph\not\in\expander}
\prob\{\graph\} = \prob\{\graph\not\in\expn\} \nonumber\\
& \stackrel{(\ref{eqn:probofexpansion})}{=} o_n(1). \label{eqn:secondterm}
\end{align}
From (\ref{eqn:firstterm}) and (\ref{eqn:secondterm}) we conclude that for
$n\geq n(\theta)$,
\begin{align*}
&\frac{1}{n} \bE[\ntouched(\graph,\errors,\iter)] \\
 \leq & \sum_{\witness:|\witness| \leq {\theta}n}
\sum_{\graph: \graph\in\expn} \prob\{\graph\} 
 \prob\{\errorset_{\graph, \witness}\} \Big(\alpha \frac{\dl}{\dr}  +
o_n(1)\Big)\\
& + \frac{\dl }{6c(\dr-1)\dr} \alpha \\
\leq & \left(\frac{\dl}{\dr}+\frac{\dl}{6c(\dr-1)\dr}\right) \alpha + o_n(1).
\end{align*}
If we now let $n$ tend to infinity then we get 
\begin{align*}
\lim_{n\to\infty}\limsup_{\ell\to\infty}\bE[P_b^{\LGalBsmall}(\graph,\epsilon,\ell)] 
\leq & \lim_{n\to\infty}\frac{1}{n}\bE[M(\graph,\errors,\ell)] \\
\leq & \left(\frac{\dl}{\dr}+\frac{\dl}{6c(\dr-1)\dr}\right) \alpha.
\end{align*}
Since this conclusion is valid for any $0 < \alpha
\leq\alpha_{\text{max}}(\gamma)$ it follows that 
\begin{align*}
\lim_{n\to\infty}\limsup_{\ell\to\infty}\bE[P_b^{\LGalBsmall}(\graph,\epsilon,\ell)]=0.
\tag*{\qed}
\end{align*}

\subsection{Extensions}\label{sec:extensions}
\subsubsection{GalB and $\dl \geq 4$}
Note that for $\dl \geq 5$ the result is already implied by Theorem~\ref{thm:limexbit}.
For $\dl=4$ the proof is easily adapted from the one for $\dl=3$. 
The only difference lies
in the way the size of the witness is computed (Section \ref{sec:witness}) and the
analysis of the birth-death process (Section \ref{sec:backtoexpansion}).

\subsubsection{MS and BSC}
The proofs can also be extended to other decoders. For a given MP
decoder, the idea is to define an appropriate {\em linearized} version of the
decoder (LMP) and go through the whole machinery as done for GalB. 

For example, consider the MS($M$) decoder and transmission over BSC($\epsilon$).
The channel realizations are mapped to $\{\pm 1\}$. Let $M \in \naturals$, the message alphabet
is $\msgspace= \{-M,\dots,M\}$. 
For transmission of the all-one codeword, the {\em linearized}
version of the decoder (LMS($M$)) is defined as in Definition \ref{def:LGalB}:
i.e., at the check node the outgoing message is the minimum of the incoming
messages and the variable node rule is unchanged.


One can check that the LMS algorithm defined above is monotonic with respect to
the input log-likelihoods at both the variable and check nodes and the number of
errors in the MS decoder can be upper bounded by the errors of the LMS decoder.

\blemma[MS($M$) Decoder, BSC and $\dl\geq3$]\label{lem:LMS}
Consider ($\dl,\dr$) ensemble and transmission over BSC($\epsilon$). Let
$\epsilon^{\LMSsmall}$ be the channel parameter below which 
$p^{(\infty)}_{\{M\}} = 1$. If $\epsilon < \epsilon^{\LMSsmall}$, then
\begin{align*}
\lim_{n\to\infty}\limsup_{\ell\to\infty}\bE[P_b^{\MSsmall}(\graph,\epsilon,\ell)]=0
\end{align*}
\elemma
\bex[LMS($2$) and BSC]
Consider communication using $\ldpc{(3,6)}$ code over BSC($\epsilon$) and decoding using MS($2$)
algorithm. For this setup, the DE threshold is $0.063$. The linearized decoder of
this algorithm has $p^{(\infty)}_{\{2\}} = 1$ for $\epsilon < 0.031$.
Therefore from the Lemma~\ref{lem:LMS} the limits can be exchanged for this
$\epsilon$.
\eex

The proof follows by showing results similar to Lemma \ref{lem:sizeofwitness}
and 
\ref{lem:upperbound}. Here we give a brief explanation for adapting the proof to
the case of $M=2$ and $\dl=3$. For a given $p >0$, we first perform $\ell(p)$ iterations such
that $p^{(\ell)}_{\{-M,\dots,M-1\}} \leq p$. We start the marking process from all the
edges with messages in $\{-M,\dots,M-1\}$ and their witness. In this case the
witness consists of edges which send messages $\{-M,\dots,M-1\}$.

To show that the size of the witness is going to zero, consider the DE equations similar to those in Appendix
\ref{appen:sizeofwitness}. Let $p_\ell^{\mu}(x)$ denote a polynomial with
non-negative coefficients where the coefficient in front of $x^i$ denotes the
probability that the message emitted by a variable node at iteration $\ell$ is
$\mu$ and that the witness (of depth $\ell$) for this edge has size $i$. Let
$q_{\ell}^{\mu}(x)$ denote the equivalent quantity for messages emitted at check
nodes. Then the DE equations for this augmented system are given by: 
\begin{align*}
p_1^{-1}(x)& = \epsilon x, \;\;\; p^{+1}_{1}(x) = \bar\epsilon x, \\
p_{\ell}^{+1}(x)& = \epsilon x( (q_{\ell-1}^{+1}(x))^2 
+2q_{\ell-1}^{+2}(1) q_{\ell-1}^{0}(x) ) + \\
&  \bar{\epsilon}x(2 q_{\ell-1}^{+2}(1) q_{\ell-1}^{-2}(x) + 2
q_{\ell-1}^{+1}(x) q_{\ell-1}^{-1}(x) + (q_{\ell-1}^{0}(x))^2),\\
p_{\ell}^{0} (x)&= \epsilon x (2 q_{\ell-1}^{+2}(1)q_{\ell-1}^{-1}(x) + 2
q_{\ell-1}^{+1}(x)q_{\ell-1}^{0}) + \\
& \bar{\epsilon} x (2 q_{\ell-1}^{+1}(x) q_{\ell-1}^{-2}(x) + 2 q_{\ell-1}^{0}(x)
q_{\ell-1}^{-1}(x)),\\
p_{\ell}^{-1}(x)& = \bar{\epsilon} x((q_{\ell-1}^{-1}(x))^2 + 2 q_{\ell-1}^{-2}(x) q_{\ell-1}^{0}(x) ) + \\
& {\epsilon}x(2 q_{\ell-1}^{+2}(1) q_{\ell-1}^{-2}(x) + 2
q_{\ell-1}^{+1}(x) q_{\ell-1}^{-1}(x) + (q_{\ell-1}^{0}(x))^2), \\
p_{\ell}^{-2}(x) &= \epsilon x2 (q_{\ell-1}^{-2}(x)(q_{\ell-1}^{+1}(x)
+q_{\ell-1}^{0}(x) +q_{\ell-1}^{-1}(x) ))\\
& + \epsilon x (2 q_{\ell-1}^{0}(x) q_{\ell-1}^{-1}(x) +(q_{\ell-1}^{-1}(x))^2
(q_{\ell-1}^{-2}(x)) ^2)\\
& + \bar{\epsilon} x (2q_{\ell-1}^{-1}(x) q_{\ell-1}^{-2}(x) +
(q_{\ell-1}^{-2}(x))^2 ),\\
q_{\ell}^{\mu}(x)& = 
\frac{p_{\ell-1}^{\mu}(x)}{p_{\ell-1}^\mu(1)}((1-\sum_{i=-M}^{\mu-1}p_{\ell-1}^i)^{\dr-1}-(1-\sum_{i=-M}^{\mu}p_{\ell-1}^i)^{\dr-1})
\end{align*}
Using the hypothesis $p_{\{M\}}^{(\infty)}=1$ and doing a similar
analysis as in Appendix \ref{appen:sizeofwitness} we can show
that the size of the witness behaves as $o_{\ell}(1)$. 
In the corresponding birth-death process we have to keep track of the size of the
set of edges with messages in $\{-M,\dots,M-1\}$. 

Similar results can be obtained for BP($M$) decoder, and channels with continuous
outputs. But the analysis of these decoders is more complicated because we have to deal
with densities of messages.

\subsubsection{MS$(M)$ and continuous channel}
Consider transmission through BMS channels with bounded output log-likelihoods
and decoding using MS($M$) decoder. For this setup it is tempting to conjecture
that the proofs
can be extended using FKG inequalities for continuous lattices \cite{Pre74}.

\section{Conclusion}
We have shown two approaches for solving the problem of limit
exchange below the DE threshold. The first one, based solely on the
expansion property of the graph, helps in proving the result for a
large class of MP decoders but only if the degree is relatively
large.  To prove the result for smaller degrees one has to include
the role of channel realizations. The second approach accomplishes this
in some cases.
In this paper we only considered channel parameters below the DE
threshold.  But the regime above this threshold is equally interesting.
One important application of proving the exchange
of limits in this regime is the finite-length analysis via a scaling
approach \cite{AMRU03} since the computation of the scaling parameters heavily
depends on the fact that this exchange is permissible.

\section*{Acknowledgment}
We would like to thank A. Montanari for suggesting to directly apply
the FKG inequalities in the proof of Lemma~\ref{lem:randomization1}
instead of the original more elaborate construction.  The work
presented in this paper is partially supported by the National
Competence Center in Research on Mobile Information and Communication
Systems (NCCR-MICS), a center supported by the Swiss National Science
Foundation under grant number 5005-67322.

\appendix

\subsection{Expansion Argument For Block Error Probability} 
The following theorem is a modified version of a theorem by Burshtein and Miller \cite{BuM00}.
\btheo[Expansion]\label{burshtein}
Consider an $(\dl,\dr,\alpha,\gamma)$ left expander. 
Assume that $0 \leq \beta \leq 1$ such that $\beta (l-1) \in \naturals$ and
that $\beta \frac{\dl-1}{\dl} \leq 2\gamma -1$. 
If at some iteration $\iter$ the number of bad variable nodes 
is less than $\frac{\alpha}{\dl\dr}n$ then the MP algorithm will decode successfully.
\etheo
\begin{proof}
Let $\bad_{\iter}$ denote the bad set in iteration $\iter$. We claim that
\begin{align}
\gamma l \vert \bad_{\iter} \cup \bad_{\iter+1} \vert 
& \stackrel{(i)}{\leq} 
\vert \nb(\bad_{\iter} \cup \bad_{\iter+1}) \vert  \nonumber \\
& \stackrel{(ii)}{\leq} 
\vert \nb(\bad_{\iter}) \vert + \beta (l-1) \vert \bad_{\iter+1} \backslash \bad_{\iter}\vert.
\label{eqn1}
\end{align}
Step (ii) follows from the fact that each variable in 
$\bad_{\iter+1} \backslash \bad_{\iter}$ must be connected to at least $l-\beta(l-1)$
checks in the set $\nb(\bad_{\iter})$ since otherwise this variable will be good
and wont be in $\bad_{\iter+1}$. Therefore the number of edges coming out
of $\bad_{\iter+1} \backslash \bad_{\iter}$ that are not connecting to $\nb(\bad_{\iter})$
is at most $\beta (l-1) \vert\bad_{\iter+1} \backslash \bad_{\iter} \vert$. 
Thus the number of neighbors of $\bad_{\iter+1} \backslash \bad_{\iter}$ that are not already
neighbors of $\bad_{\iter}$ is at most $\beta (l-1) \vert\bad_{\iter+1} \backslash \bad_{\iter} \vert$.

Consider now step (i). This step follows in a straightforward 
fashion from the expansion property since by assumption $|\bad_{\iter}| \leq \frac{\alpha}{\dl \dr} n$ so
that $|\bad_{\iter} \cup \bad_{\iter+1}| < \alpha n$.

Let $T$ be the set of check nodes that are connected to $\bad_{\iter} \cap \bad_{\iter+1}$ but not connected to $\bad_{\iter} \backslash\bad_{\iter+1}$. Suppose an edge from a check node in $T$ is carrying a bad message. Then this check must be connected to one more variable in $\bad_{\iter}\cap \bad_{\iter+1}$ because it is not connected to $\bad_{\iter} \backslash \bad_{\iter+1}$ and thus cannot get a bad message from $\bad_{\iter} \backslash \bad_{\iter+1}$. For each variable in 
$\bad_{\iter}\cap \bad_{\iter+1}$, at least $l-\beta(l-1)$ edges must be bad messages and hence it can connect to at most 
$(l-\beta(l-1))/2 +\beta(l-1)=l/2+\beta(l-1)/2$ check nodes. Therefore we have,
\begin{align}\label{eqn2}
\vert \nb(\bad_{\iter})\vert &\leq l\vert \bad_{\iter} \backslash \bad_{\iter+1}\vert + 
\vert T \vert, \nonumber\\
|\nb(\bad_{\iter})| & \leq l\vert \bad_{\iter}\backslash \bad_{\iter+1}\vert + 
\frac{1+\beta \frac{l-1}{l}}{2} l \vert \bad_{\iter} \cap \bad_{\iter+1}\vert.
\end{align}
Using equations $(\ref{eqn1})$ and $(\ref{eqn2})$, we get
\begin{align*}
\gamma l \vert \bad_{\iter+1} \cup \bad_{\iter}\vert \leq &  
l\vert \bad_{\iter} \backslash \bad_{\iter+1}\vert +  \frac{1+\beta \frac{l-1}{l}}{2} l \vert \bad_{\iter+1}\cap \bad_{\iter}\vert  \\
& +\beta (l-1) \vert \bad_{\iter+1}\backslash\bad_{\iter}\vert \nonumber \\
\gamma \vert \bad_{\iter+1} \cap \bad_{\iter}\vert & + \gamma \vert \bad_{\iter} 
\backslash \bad_{\iter+1}\vert + \gamma \vert \bad_{\iter+1} 
\backslash \bad_{\iter}\vert  \nonumber\\
\leq  &   \vert \bad_{\iter} \backslash \bad_{\iter+1}\vert + 
\frac{1+\beta \frac{l-1}{l}}{2} \vert \bad_{\iter+1}\cap \bad_{\iter}\vert + \\
& \beta \frac{l-1}{l} \vert \bad_{\iter+1}\backslash\bad_{\iter}\vert \nonumber \\
\vert \bad_{\iter+1}\backslash\bad_{\iter}\vert \leq &  \frac{(1-\gamma)}{\gamma -
\beta \frac{l-1}{l}}\vert \bad_{\iter}\backslash \bad_{\iter+1}\vert +\nonumber \\ 
&\frac{1+\beta \frac{l-1}{l}  -2\gamma}{2(\gamma -\beta \frac{l-1}{l})}\vert \bad_{\iter} \cap \bad_{\iter+1}\vert 
\end{align*}
The coefficient of the first term in RHS is less than $1$ and the coefficient of
the second term is negative and hence
$\vert\bad_{\iter+1}\backslash\bad_{\iter}\vert <
\vert\bad_{\iter}\backslash\bad_{\iter+1}\vert$
\end{proof}

\subsection{Size of Witness}\label{appen:sizeofwitness}
\noindent{\em Proof of Lemma~\ref{lem:sizeofwitness}.}
Let $\graph$ be a graph and let $\errors$ be the noise realization.
Assume that we perform $\ell$ iterations.
Let $\witness_{\edge}(\graph,\errors,\ell)$ denote the witness of edge $\edge$.
Then
\begin{align*}
\bE[|\witness(\graph,\errors,\ell)|] 
\leq \sum_{i=1}^{\dl n} 
\bE[|\witness_{\edge_i}(\graph,\errors,\ell)|] = 
n\dl \bE[|\witness_{\edge_1}(\graph,\errors,\ell)|].
\end{align*}
It remains to compute the expected size of the witness for the limit of $n$ tending
to infinity and a fixed $\ell$. This can be accomplished by DE.

Let $\de{\iter}$ denote the probability of an edge being in error
according to DE.  Let $p_{\iter}(x)$ denote a
polynomial with non-negative coefficients where the coefficient in
front of $x^i$ denotes the probability that the message emitted by
a variable node at iteration $\iter$ is bad and that the witness
(of depth $\ell$) for this edge has size $i$ ($i$ variable nodes). Let $q_{\iter}(x)$
denote the equivalent quantity for messages emitted at check nodes.
The DE equations for this augmented system are:
\begin{align*}
p_1(x) & = \epsilon x, \\
p_{\ell}(x) & = \epsilon (2-q_{\iter}(1)) q_{\iter}(x) x + \bar{\epsilon} q_{\iter}(x)^2 x,  \\
q_{\ell}(x) & = \frac{p_{\ell-1}(x)}{p_{\ell-1}(1)} (1-(1-p_{\ell-1}(1))^{\dr-1}).
\end{align*}
The initialization $p_1(x)  = \epsilon x$ reflects the fact that
with probability $\epsilon$ a variable-to-check message is in error
in iteration $1$ and that its associated witness of depth $1$
consists only of the attached variable (hence the $x$).

The recursion for $q_{\ell}(x)$ is also straightforward.  With
probability $1-(1-p_{\ell-1}(1))^{\dr-1}$ at least one of the $\dr-1$
incoming messages at a check node is bad, and in this case the
distribution of the size of the attached witness is 
$\frac{p_{\ell-1}(x)}{p_{\ell-1}(1)}$.

Let us now look at the recursion for $p_{\ell}(x)$.  There are three
contributions: (i) Suppose that the variable has a bad received
value and that exactly one of the incoming edges is bad; this happens
with probability $\epsilon 2(1-q_{\iter}(1)) q_{\iter}(1)$ and in
this case the distribution of the size of the witness attached to
this edge is $\frac{q_{\iter}(x) x}{q_{\iter}(1)}$, where the extra
$x$ accounts for the attached variable node.  (ii) Suppose that the
variable has a bad received value and that both incoming edges are
bad; this happens with probability $\epsilon q_{\iter}(1)^2$, and
in this case the distribution of the size of the witness attached
to this edge is $\frac{q_{\iter}(x) x}{q_{\iter}(1)}$.  (iii)
Finally, suppose that the variable has a good received value and
that both the incoming edges are bad; this happens with probability
$\bar{\epsilon} q_{\iter}(1)^2$ and in this case the distribution
of the size of the witness attached to this edge is $\frac{q_{\iter}(x)^2
x}{q_{\iter}(1)^2}$.

Note that we get standard DE by 
setting $x=1$, i.e., we have $x_{\iter} = p_{\iter}(1)$. 
We want to show that $p_{\ell}'(1)$ 
(this is the expected size of the witness in the limit of infinite blocklengths) 
converges to zero as a function of $\ell$.

The augmented DE equation is difficult to handle.
So let us first write down a scalar version that tracks the expected value.
Define $\beta_{\ell}=\frac{(1-(1-p_{\ell}(1))^{\dr-1})}{p_{\ell}(1)}$.
Then we get
\begin{align*}
p_{\ell}(x) & = \epsilon (2-q_\ell(1)) \beta_{\ell-1} p_{\iter-1}(x) x + 
\bar{\epsilon} \beta_{\ell-1}^2 p_{\iter-1}(x)^2 x.
\end{align*}
Differentiate both sides with respect to $x$. This gives 
\begin{align*}
p_\ell'(x)  = & \epsilon \beta_{\ell-1} (2-q_\ell(1))(p'_{\ell-1}(x) x
+p_{\ell-1}(x))\\
& + \bar{\epsilon}\beta_{\ell-1}^2 (p_{\ell-1}(x))^2 + \bar{\epsilon}
\beta^2_{\iter -1} 2
p_{\ell-1}(x) p'_{\ell-1}(x)x.
\end{align*}
Now substitute $x=1$. Recall that  $x_{\iter} = p_{\iter}(1)$ and define $p_\ell =
p'_{\ell}(1)$. Further, bound $2-q_{\ell}(1)$ by $2$ and $\beta_{\ell}$ by
$(\dr-1)$. This gives the inequality
\begin{align*}
p_{\ell} \leq & 2\epsilon (\dr-1) p_{\ell-1} + 2\epsilon (\dr -1) x_{\ell-1} \\
&+\bar{\epsilon}(\dr-1)^2x_{\ell-1}^2 + 2 \bar{\epsilon}(\dr-1)^2
x_{\ell-1}p_{\ell-1}.
\end{align*}
We claim that $\ell \de{\ell} \leq p_\ell$. This is true since
$x_{\ell}$ is the probability of a bad message, whereas $p_{\ell}$ is
the expected size of the witness and the witness size is always at least $\ell$
if the message is bad.
Therefore,
\begin{align*}
\frac{p_{\ell}}{p_{\ell-1}} \leq & 2\epsilon (\dr-1) + 2\epsilon (\dr -1)
\frac{x_{\ell-1}}{p_{\ell-1}} \\
&+\bar{\epsilon}(\dr-1)^2\frac{x_{\ell-1}^2}{p_{\ell-1}} + 2 \bar{\epsilon}(\dr-1)^2
x_{\ell-1}\\
 \leq & 2\epsilon(\dr-1) + 2\epsilon\frac{(\dr-1)}{\ell} + 3\bar{\epsilon}(\dr-1)^2x_{\ell-1}.
\end{align*}
Now note that $x_{\ell}$ tends to zero since $\epsilon < \epsilon^{\LGalBsmall}$. Therefore, if $2 \epsilon (\dr-1) < 1$
then $p_{\ell}/p_{\ell-1}<1$ for $\ell$ sufficiently large. The stability
condition implies $\epsilon^{\LGalBsmall} < \frac{1}{2(\dr-1)}$.
Therefore, for $\epsilon < \epsilon^{\LGalBsmall}$, $p_{\ell}$ tends to zero exponentially fast
for increasing $\ell$.
\qed

\subsection{Randomization}\label{apen:randomization}
\noindent{\em Proof of Lemma~\ref{lem:randomization2}.}
We have
\begin{align*}
& 
\bE[\ntouched(\graph, \errors, \ell)] \\
&=  
\sum_{\witness} \bE[\ntouched(\graph, \errors, \witness)
\indicator{\witness(\graph, \errors, \ell) = \witness}] \\
&=  
\sum_{\witness,\graph} \prob\{\graph\}
\bE_{\errors} 
[
\ntouched(\graph, \errors, \witness)
\indicator{\witness(\graph, \errors, \ell) = \witness } 
] \\
&=  
\sum_{\witness,\graph} \prob\{\graph\}
\bE_{\errors} 
[
\ntouched(\graph, \errors, \witness)
\indicator{\errors \in \errorset_{\graph, \witness}}].
\end{align*}
For all $\errors \in \errorset_{\graph,\witness}$, the channel values on
$\witness$ are fixed to those appearing in the witness which is also denoted
by $\witness$. Recall that $\errorset'_{\graph,\witness}$ is the projection of
$\errorset_{\graph,\witness}$ on $\graph\backslash \witness$ and $\errors' \in
\errorset'_{\graph,\witness}$. The above expectation is equivalent to
\begin{align*}
\bE_{\errors}&
[ \ntouched(\graph, (\witness,\errors'),
\witness)\indicator{(\witness,\errors')\in\errorset_{\graph,\witness}}] =\\
&\prob(\witness)\bE_{\errors'} [ \ntouched(\graph, (\witness,\errors'),
\witness)\indicator{\errors'\in\errorset'_{\graph,\witness}}],
\end{align*}
where $\prob(\witness)$ is the probability of the channel values on $\witness$.
This implies $\prob(\witness)\prob(\errorset'_{\graph,\witness}) =
\prob(\errorset_{\graph,\witness})$. 
Using (\ref{eq:randomization}) we bound 
\begin{align*}
\bE_{\errors'}& [ \ntouched(\graph, (\witness,\errors'),
\witness)\indicator{\errors'\in\errorset'_{\graph,\witness}}] \\
& \leq  \prob(\errorset'_{\graph,\witness})
\bE_{\errors'} [ \ntouched(\graph, (\witness,\errors'),
\witness)].
\end{align*}
Therefore,
\begin{align*}
&\bE[\ntouched(\graph, \errors, \ell)] \\
& \leq
\sum_{\witness,\graph} \prob\{\graph\}
\prob\{\errorset_{\graph, \witness}\}
\bE_{\errors'}[\ntouched(\graph, (\witness, \errors'), \witness)]  \\
&\leq 
\sum_{\witness:|\witness| \leq {\theta}n,\graph} \prob\{\graph\}
\prob\{\errorset_{\graph, \witness}\}
\bE_{\errors'} [\ntouched(\graph, (\witness, \errors'), \witness)] + \\
&\;\;\;
\sum_{\witness :|\witness| \geq {\theta}n,\graph} \prob\{\graph\}
\prob\{\errorset_{\graph, \witness}\}
\bE_{\errors'}
[\ntouched(\graph, (\witness, \errors'), \witness)].
\end{align*}
Consider the second term in the last line. Bound the expectation by $n$. This yields 
\begin{align*}
\sum_{\witness :|\witness| \geq {\theta}n,\graph}&
\prob\{\graph\}
\prob\{\errorset_{\graph, \witness}\} n.
\end{align*}
If $\witness \not\subseteq \graph$, then $\errorset_{\graph,\witness}$ is empty.
Therefore the above bound is equivalent to
\begin{align*}
&n\sum_{\witness:|\witness| \geq {\theta}n}
\bE[\indicator{\witness \subseteq \graph}\indicator{\errors\in
\errorset_{\graph,\witness}}]\\
&= n \sum_{\witness:|\witness| \geq {\theta}n}
\bE[\indicator{\witness(\graph,\errors,\iter)=\witness}]\\
& = n\prob\{|\witness(\graph,\errors,\iter)| \geq {\theta} n\}.
\end{align*}
By assumption, $\bE[|\witness(\graph,\errors, \ell)|] \leq \theta^2 n$.
The Markov inequality therefore shows that
\begin{align*}
\prob\{|\witness(\graph,\errors, \ell)| \geq {\theta} n\} \leq {\theta}.
\tag*{\qed}
\end{align*}

\subsection{FKG Inequality}\label{appen:FKG}
Consider the Hamming space $\{0,1\}^n$. 
For $x, y \in \{0, 1\}^n$ define the following partial order:
$x \leq y$ iff $x_i \leq y_i$ for all $i$. Define $x_{\leq}$ as 
\begin{align}
x_{\leq} = \{y: y\in\{0,1\}^n,\;y\leq x\},
\end{align}
and
$x\vee y$ and $x \wedge y$ as
\begin{align}
(x\vee y)_i = \left\{
\begin{array}{cl}
0 & \text{if } x_i = y_i = 0,\\
1 & \text{else},
\end{array}\right.
\end{align}
\begin{align}
(x\wedge y)_i = \left\{
\begin{array}{cl}
1 & \text{if } x_i = y_i = 1,\\
0 & \text{else}.
\end{array}\right.
\end{align}

We say that a function $f: \{0, 1\}^n \rightarrow \reals$ is 
monotonically increasing (decreasing) if $f(x) \geq f(y)$ whenever $x \geq y$ ($x \leq y$).
\begin{lemma}[FKG Inequality -- \cite{FoKG71}]
\label{lem:fkg}
Let $P\{ \cdot \}$ be a probability measure on $\{0,1\}^n$ such that
\begin{align*}
P\{x\}P\{y\} \leq P\{x \vee y\} P\{x\wedge y\}.
\end{align*}
Let $f$ and $g$ be real-valued non-negative functions on $\{0, 1\}^n$.
If $f$ and $g$ are either both monotonically increasing or both decreasing then
\begin{align*}
\bE[f(x)g(y)] \geq \bE[f(x)]\bE[g(y)].
\end{align*}
\end{lemma}

\subsection{Birth and Death Process}\label{appen:birthdeath}
Consider the following birth and death process. We start with $X_0 = a > 0$. 
At step $t$, $t \in \naturals$, if $X_{t-1} < 1$ then we stop the process
and define $X_{t'}= X_{t'-1}$ for $t' > t$.
Otherwise we decrease $X_{t-1}$ by $1$ and add $Y_t$, where the sequence
$\{Y_t\}_{t \geq 1}$ is iid.
In this way, as long as $X_{t-1} \geq 1$,
\begin{align*}
X_{t}=X_{t-1}-1+Y_{t}.
\end{align*}
This process is equivalent to the standard birth and death process if $Y_t$ takes 
non-negative integer values. In this case, the step described above corresponds to  
choosing a member of the population which then creates $Y_t$ off-springs and dies.

Let $T$ denote the stopping time, i.e., $T=\min\{t: X_t < 1 \}$.
 
\begin{lemma}[Birth-Death]\label{apenlem:birthdeath}
Fix $p\in (0, 1]$ and $0<\mu < 1$.
Consider a birth and death process with $X_0=a \in \naturals$ and
\begin{align*}
Y_i & =
\begin{cases}
\frac{\mu}{p}, & \text{with probability $p$},  \\
0, & \text{with probability $1-p$},
\end{cases}
\end{align*}
so that $\bE[Y_i]=\mu$. 
Then, for $\beta a \in \naturals$,
\begin{align*}
\prob\{T > \beta a \}  & \leq  e^{-ac(p,\mu,\beta)}
\end{align*}
where $c(p,\mu,\beta) > 0$ for $\beta > \frac{1}{1-\mu}$.
\end{lemma}
\begin{proof}
Let $b = \beta a$.
Note that 
\begin{align*}
\prob\{T > b \} \leq \prob \{X_b \geq 1 \}\leq \prob\{X_b \geq 0\}.
\end{align*}
Let $\tilde{Y}_t = Y_t - 1$.
We have
\begin{align*}
X_b = X_{b-1} + \tilde{Y}_b = X_{b-2} +\tilde{Y}_{b-1}+\tilde{Y}_b = 
a + \sum_{i=1}^{b} \tilde{Y}_i.
\end{align*}
Therefore,
\begin{align*}
\prob \{T > b \}
& \leq \prob \Bigl\{\sum_{i=1}^b \tilde{Y}_i \geq -a \Bigr\}
\stackrel{s>0}{=} \prob \bigr\{e^{s\sum_{i=1}^{b} \tilde{Y}_i} \geq e^{-as} \bigr\}\\
& \stackrel{\text{Markov}}{\leq} e^{as} \bE[e^{s\tilde{Y}}]^b 
= e^{as} \Bigl((1-p) e^{-s} +p e^{(\frac{\mu}{p}-1) s} \Bigr)^b.
\end{align*}
First consider the case $\mu \geq p$. Set $s = \frac{p}{\mu}\ln\frac{(\beta -1)(1-p)}{p+\beta(\mu-p)}$, which is strictly positive
since $\mu\geq p$ and $\beta > \frac{1}{1-\mu}$. Set $\beta=\frac{1}{1-\mu-\xi}$,
where $\xi>0$. With this choice we get
\begin{align*}
\prob \{T > b \} &\leq \Bigl[\frac{\mu(1-p)}{\mu(1-p)-\xi p} \Bigl( \frac{\mu(1-p)-\xi p}{\mu(1-p)+\xi (1-p)} \Bigr)^{\frac{p (\mu+\xi)}{\mu}}\Bigr]^b.
\end{align*}
For $\xi=0$ the terms inside the square brackets is $1$.
If we take the derivative of the expression inside the square brackets wrt to
$\xi$ we get
\begin{align*}
\frac{-p}{\mu+\xi} \Bigl(\frac{ (\mu +\xi) (1-p)}{\mu(1-p)-p \xi}\Bigr)^{1-\frac{p(\mu+\xi)}{\mu}}
\log \frac{(\mu +\xi) (1-p)}{\mu(1-p)-\xi p}.
\end{align*}
For $\xi>0$ and $\mu>p$ this is strictly negative which proves our claim.

Now consider the case $\mu < p$. 
For $\frac{1}{1-\mu}< \beta < \frac{p}{p-\mu}$ the above still applies.
For $\beta \geq \frac{p}{p-\mu}$, the probability is
$0$. This is because in each step we can add at most $\frac{\mu}{p} - 1$.
Therefore, for $t \geq \frac{p}{p-\mu}a+1$, $X_t \leq a +
(\frac{p}{p-\mu}a+1)(\frac{\mu}{p} - 1) < 0$. 
\end{proof}

\subsection{Concentration}
\btheo[Concentration Theorem \protect{\cite{RiU08}[p. 222]}] 
\label{the:concentration} Let $\graph$, chosen uniformly at random from
LDPC$(n, \ledge, \redge)$, be used for transmission over a BMS$(\epsilon)$ channel.
Assume that 
the decoder performs $\iter$ rounds of message-passing
decoding and let $P_b^{\MPsmall}(\graph, \epsilon, \iter)$ 
denote the resulting bit error probability. Then, for
any given $\delta>0$, there exists an $\alpha >0$,
$\alpha=\alpha(\ledge, \redge, \delta)$, such that
\[
\prob\{|P_b^{\MPsmall}(\graph, \epsilon, \iter)-
\expectation_{\ldpc (n,\ledge,\redge)}\left[P_b^{\MPsmall}(\graph, \epsilon, \iter) \right]|
> \delta \} \leq e^{-\alpha n}.
\]
\etheo

\bibliographystyle{IEEEtran} 
\bibliography{lth,lthpub}

\end{document}